\documentclass[titlepage,12pt]{article}
\usepackage{amssymb,amsmath,color,graphics,amscd,epsf,indentfirst,amsfonts}
\usepackage{enumerate}
\usepackage{epsfig}
\usepackage[hypertex]{hyperref}
\usepackage{nicefrac}
\usepackage{scalefnt}
\usepackage[titles]{tocloft}
\usepackage{sectsty}
\allsectionsfont{\bf \scalefont{.7} \selectfont}
\subsectionfont{\bf \scalefont{.85} \it \selectfont}
\subsubsectionfont{\bf \scalefont{1} \it \selectfont}
\usepackage[T1]{fontenc}
\usepackage{lmodern}
\usepackage{bibspacing}

\def\blfootnote{\xdef\@thefnmark{}\@footnotetext}

\long\def\symbolfootnote[#1]#2{\begingroup%
\def\thefootnote{\fnsymbol{footnote}}\footnote[#1]{#2}\endgroup}

\makeatletter
\renewcommand{\@dotsep}{4.5}
\makeatother

\def\be{\begin{equation}}
\def\ee{\end{equation}}

\makeatletter
\def\@seccntformat#1{\csname the#1\endcsname.\quad}
\makeatother

\def\clock{{\count0=\time
           \divide\count0 60
           \ifnum\count0<10 0\fi\the\count0
           \multiply\count0 -60 \advance\count0 \time
           :\ifnum\count0<10 0\fi \the\count0
         }}
\newcommand{\timestamp}{{\small\vbox{\hbox{\tt\jobname.tex}
\hbox{\the\day/\the\month/\the\year, \clock}}}}

\def\AA{{\cal A}}
\def\BB{{\cal B}}
\def\CC{{\cal C}}

\def\KK{{\cal K}}
\def\LL{{\cal L}}

\def\NN{{\cal N}}
\def\OO{{\cal O}}

\def\SS{{\cal S}}
\def\TT{{\cal T}}

\def\VV{{\cal V}}

\def\d{{\partial}}

\def\ddbar{{${\rm D}\overline{\rm D}$}}
\def\csb{$\chi{\rm s B}$}

\newcommand{\beq}{\begin{equation}}
\newcommand{\eeq}{\end{equation}}
\newcommand{\ba}{\begin{array}}
\newcommand{\ea}{\end{array}}
\newcommand{\bea}{\begin{eqnarray}}
\newcommand{\eea}{\end{eqnarray}}

\newcommand{\Z}{\mathbb{Z}}
\newcommand{\C}{\mathbb{C}}
\newcommand{\R}{\mathbb{R}}

\numberwithin{equation}{section}

\begin{document}

\begin{titlepage}
\begin{flushright}
CPHT-038.0510
\end{flushright}
\vskip 3.3cm
\begin{center}
\font\titlerm=cmr10 scaled\magstep4
    \font\titlei=cmmi10 scaled\magstep4
    \font\titleis=cmmi7 scaled\magstep4
    \centerline{\titlerm
     Hairpin-Branes and Tachyon-Paperclips
      \vspace{0.4cm}}
      \centerline{\titlerm
      in Holographic Backgrounds}   
\vskip 1.5cm
{\it Vasilis Niarchos}\\
\vskip 0.7cm
\medskip
{Centre de Physique Th\'eorique, \'Ecole Polytechnique}\\
{CNRS, 91128 Palaiseau, France}\\
\medskip
{niarchos@cpht.polytechnique.fr}

\end{center}
\vskip .6cm
\centerline{\bf Abstract}

\baselineskip 20pt
%

\vskip .7cm \noindent
D-branes with a U-shaped geometry, like the D8 flavor branes in the Sakai-Sugimoto
model of QCD, are encountered frequently in holographic backgrounds. We argue that 
the commonly used DBI action is inadequate as an effective field theory description of 
these branes, because it misses a crucial component of the low-energy dynamics: a 
{\it light} complex scalar mode. Following an idea of Erkal, Kutasov and Lunin we 
elaborate on an effective description based on the {\it abelian} tachyon-DBI action 
which incorporates naturally the non-local physics of the complex scalar mode. We 
demonstrate its power in a context where an explicit worldsheet description of the 
open string dynamics exists ---hairpin-branes in the background of NS5-branes. Our 
results are relevant for the holographic description of chiral symmetry breaking and 
bare quark mass in QCD and open string tachyon condensation in curved backgrounds.

\vfill
\noindent
May, 2010
\end{titlepage}\vfill\eject

\setcounter{equation}{0}

\pagestyle{empty}
\small
\tableofcontents
\normalsize
\pagestyle{plain}
\setcounter{page}{1}

\section{Introduction}
\label{intro}

An important problem in holographic discussions of large-$N$ gauge theories is 
how to describe efficiently the properties of systems that exhibit flavor chiral symmetry
breaking (\csb). By now a variety of holographic setups that exhibit chiral symmetry breaking
is known. In the quenched approximation the flavor degrees of freedom are realized by a 
set of D-branes that extend in the radial direction of the holographic background. In some of 
these setups, the breaking of the flavor chiral symmetry occurs as a D-brane reconnection 
process.\footnote{The situation is different, for instance, in the D4-D6 system of Ref.\ 
\cite{Kruczenski:2003uq}.} To be concrete, let us consider a few well known examples.

The holographic physics of chiral symmetry breaking was discussed already in the early 
work \cite{Karch:2002sh}. One of the examples in that paper, based on a D3-D7-D7$'$
system in type IIB string theory, realizes an $\NN=2$ gauge theory that can be viewed as 
$\NN=4$ super-Yang-Mills (SYM) theory with $2N_f$ extra hypermultiplets. The $N_f$ D7 and 
$N_f$ D7$'$ branes, which are oriented in orthogonal directions ---the D7's are placed at 
$z=X^8+iX^9=0$ and the D7$'$'s at $w=X^6+iX^7=0$---, carry a non-chiral $U(N_f)\times U(N_f)$ 
flavor symmetry group. In gauge theory this group is broken explicitly to its diagonal with a 
mass deformation for the fundamental hypermultiplets. In the dual AdS$_5\times$S$^5$ 
background the D7-D7$'$ pairs reconnect to wrap a smooth curve of the form 
$zw=\varepsilon\neq 0$.

The Sakai-Sugimoto model of large-$N$ QCD is another characteristic example 
\cite{Sakai:2004cn,Sakai:2005yt}. This model is based on a D4-D8-$\overline {\rm D8}$ 
system where a $U(N_f)\times U(N_f)$ flavor symmetry group is realized on a stack of 
$N_f$ D8-$\overline{\rm D8}$ pairs. The antiparallel D8 and anti-D8 branes in each pair are 
separated in a transverse direction. At large 't Hooft coupling, where the supergravity 
description of the D4-branes is appropriate, the D8-$\overline{\rm D8}$ pairs reconnect 
to $N_f$ D8-branes with a U-shaped geometry. This feature reproduces naturally the 
spontaneous chiral symmetry breaking that occurs in the dual QCD-like gauge theory.

Qualitatively similar open string physics takes place in a non-supersymmetric Little String 
Theory (LST) version of the Sakai-Sugimoto model (for a review of LST see \cite{Aharony:1999ks}). 
Replacing the D4-branes by $k$ NS5-branes we obtain a stack of $\ell$ D$p$-$\overline{{\rm D}p}$ 
pairs ($1\leq p\leq 5$) intersecting orthogonally the NS5-branes. In the gravitational dual of this 
system the D$p$-$\overline{{\rm D}p}$ pairs reconnect to $\ell$ D$p$-branes with a U-shaped 
embedding in the curved near-horizon background of the NS5 branes. Because of their U shape, 
these branes are frequently called hairpin-branes.

Each of these setups requires a detailed understanding of the open string physics on the reconnected
hairpin flavor branes. The resolution of fundamental questions about the flavor sector of the strongly 
coupled dual gauge theory depends on this understanding, $e.g.$ questions about the order parameter 
of \csb, bare quark masses, the structure of the mesonic spectra, $etc$. Lacking an explicit solution
of string theory, in many cases we rely heavily on effective field theory descriptions. The standard 
description of low-energy dynamics in open string theory is the Dirac-Born-Infeld (DBI) action. In 
this paper we will argue that this description is inadequate for the kind of D-brane systems that 
we discussed above.  

In the first example of the D3-D7-D7$'$ system it was already pointed out in \cite{Karch:2002sh}
that the mass deformation that breaks the flavor chiral symmetry explicitly corresponds to giving a 
vacuum expectation value (vev) to a bifundamental hypermultiplet. This hypermultiplet arises 
as a massless mode in the NS$-$ sector of an open string that is localized in the six dimensional 
intersection of the orthogonal D7 and D7$'$-branes. An effective action that includes the non-linear 
completion of the six dimensional hypermultiplet is needed to describe the full dynamics of this system
\cite{Karch:2002sh}.

A similar effect occurs in the second example, the Sakai-Sugimoto model. 
It is clear holographically that the chiral symmetry breaking that takes place in this case 
should be attributed again to the vev of a complex scalar field that belongs in the 
bifundamental representation of the $U(N_f)\times U(N_f)$ flavor symmetry group. This 
field arises as a ground state in the NS$-$ sector of open strings that stretch between the 
D8 and anti-D8 branes. In flat space this mode is tachyonic when the separation of the
D8 and anti-D8 branes $L$ is comparable to the string scale $\ell_s$ and highly massive 
when $L\gg \ell_s$. Because of this, one may be tempted to conclude that there is a similar
regime in the Sakai-Sugimoto model where the ground state in the NS$-$ sector of the 
D8-$\overline{{\rm D}8}$ strings is also highly massive  and therefore irrelevant for the infrared 
physics. We will argue that this intuition from flat space is in fact misleading and that this 
mode can be light in curved backgrounds even when $L\gg \ell_s$. In holographic setups 
the mass formula of this mode can get a large negative contribution from the non-trivial 
profile of the wavefunction in the radial direction that cancels the large positive contribution 
related to the separation $L$. A similar picture was advocated in \cite{Antonyan:2006vw}. 
Here we set up an effective description of open string dynamics that incorporates naturally 
this effect and corroborates this picture.

In the third example, the NS5-D$p$-$\overline{{\rm D}p}$ system, we can exhibit this 
effect very explicitly by solving the open string theory on the D$p$-$\overline{{\rm D}p}$
branes with the use of $\alpha'$-exact methods of worldsheet conformal field theory (CFT). 
We can identify the NS$-$ sector mode of interest, compute its mass and see how it 
condenses on the D$p$-branes.\footnote{The automatic appearance and relevance 
of a non-vanishing vev for a normalizable mode of the bifundamental scalar was also 
emphasized in Ref.\ \cite{Fotopoulos:2005cn} in a related context---the context of the 
non-critical near-horizon background of two orthogonal NS5-branes, where probe D3 and 
D5-branes were used to engineer the four-dimensional $\NN=1$ supersymmetric QCD. 
In that case the spacefilling flavor D5-branes are T-dual to D4 hairpin-branes and the
appearance of a non-vanishing vev for the bifundamental scalar has a natural interpretation
in SQCD.}

Our purpose in this paper is to set up an effective field theory description of the low-energy 
open string dynamics on U-shaped reconnected D$p$-$\overline{{\rm D}p}$ 
systems of the type presented in the second and third examples above. Besides the
transverse scalars and gauge fields this action must also capture the dynamics 
of a complex scalar field in the bifundamental of the flavor gauge group that comes
from a long open string of the D$p$-$\overline{{\rm D}p}$ system. Following the standard
nomenclature of \ddbar\ systems we will call this mode a tachyon, but it should be 
remembered that it gives rise to a massless field in the systems that we will examine.
Finding a well motivated, efficient description of this system is currently a largely open problem.

Previous attempts \cite{Bergman:2007pm,Dhar:2008um} to incorporate this bifundamental 
mode into the analysis of the Sakai-Sugimoto model were based on a non-abelian tachyon-DBI 
(TDBI) action that has been proposed \cite{Sen:2003tm,Garousi:2004rd} to describe the 
physics of \ddbar\ systems in flat space. Unfortunately, it is unclear, already in flat space, to 
what extent  these non-abelian actions provide a good effective field theory description of 
open string physics. Moreover, it is unclear in the application of Refs.\ 
\cite{Bergman:2007pm,Dhar:2008um} how one incorporates the non-local physics of the 
bifundamental field.\footnote{This problem is absent in the phenomenological five-dimensional 
setup of Ref.\ \cite{Casero:2007ae} where the branes are coincident with the anti-branes.
For a more recent phenomenological application of the Sen action \cite{Sen:2003tm}
in coincident brane-antibrane systems see \cite{Iatrakis:2010zf}.}

In what follows, we will take a different route based on a recent proposal for a new 
effective description of the dynamics of \ddbar\ systems in flat space put forward by Erkal,
Kutasov and Lunin in Ref.\ \cite{Erkal:2009xq}. The basis of this approach is the {\it abelian}
tachyon-DBI action (we will concentrate on a single D$p$-$\overline {{\rm D}p}$ pair and set 
$\alpha'=1$) 
\begin{subequations}
\label{introaa}
\beq
\label{introaaa}
\SS=-\int d^{p+2}x\, V(T) \sqrt{-\det A}
~,
\eeq
\vspace{-.5cm}
\beq
\label{introaab}
A_{ab}=\eta_{ab}+\d_a X^I \d_b X^I+2\pi F_{ab}+\d_a T \d_b T
~,
\eeq
\vspace{-0.5cm}
\beq
\label{introaac}
V(T)=\frac{\tau_{p+1}}{\cosh (\alpha T)}
\eeq
\end{subequations}
that describes the dynamics of a {\it real} tachyon $T$ on a non-BPS D$(p+1)$-brane with tension 
$\tau_{p+1}$. The constant $\alpha$ that appears in the tachyon potential \eqref{introaac} 
is $1$ for the bosonic string and $\frac{1}{\sqrt{2}}$ for the type II string. The D$p$-$\overline{{\rm D}p}$ system arises in this description as a kink-antikink solution. The complex \ddbar\ tachyon, the 
$U(1)\times U(1)$ gauge field and the extra transverse scalars are emergent degrees of freedom. 
We will review the basic features of this formalism in section \ref{ddbar}. The formalism involves 
a particular type of solutions of the action \eqref{introaa} with a `paperclip' profile in an extended 
spacetime that includes $T$ as a fictitious extra spacetime coordinate (the second ingredient 
of the title of the present work ---`tachyon-paperclips'--- refers to this type of solutions).

In contrast to the proposed non-abelian DBI actions for the \ddbar\ tachyon, the action \eqref{introaa} 
has been derived from open string theory in a well-defined decoupling limit and is known to
describe small fluctuations around the rolling tachyon background (more precisely, around 
the `half S-brane') \cite{Kutasov:2003er,Niarchos:2004rw}.\footnote{Strictly speaking, the 
derivation of \cite{Kutasov:2003er,Niarchos:2004rw} was performed for vanishing velocities 
and gauge field strengths. The more general action \eqref{introaa} passes a number of non-trivial 
tests \cite{Sen:2003tm,Erkal:2009xq} and is believed to be the correct way to incorporate the 
transverse scalars and abelian gauge field.} In this derivation the rolling tachyon solution plays 
the same role that solutions with constant electromagnetic field and/or velocity play in the case 
of the usual DBI action. 

In the rest of this paper, we would like to treat the action \eqref{introaa} as a well motivated toy field 
theory action for open string dynamics and apply it to general closed string backgrounds. Part of 
our exercise will be to verify that it produces results consistent with knowledge from open string 
theory in curved (holographic) backgrounds (NS5-brane backgrounds in this 
work). We will assume the validity of the general form of the action \eqref{introaaa}, 
with the obvious incorporation of the induced metric in $A_{ab}$, but will keep the precise form 
of the tachyon potential free allowing in this factor the possibility of a background dependence.
We will discuss the motivation for leaving such a possibility open. In the context of the NS5-brane
system we will present a tachyon-paperclip solution that reproduces anticipated features of 
string theory but requires a modification of the potential \eqref{introaac}. 

The main lessons of this paper can be summarized as follows:
\begin{itemize}
\item[(1)] We will see explicitly in section \ref{ssNS5}, in the context of the 
NS5-D$p$-$\overline{{\rm D}p}$ system, how a complex scalar mode from a long open 
string arises in the low-lying perturbative spectrum of the open string theory on the 
U-shaped D$p$-branes. On the worldsheet, a marginal boundary interaction of this 
mode is dual to a boundary interaction that captures the geometric bending of the 
brane \cite{Lukyanov:2003nj,Hosomichi:2004ph,Kutasov:2005rr,Lukyanov:2005bf}. 
Both interactions are turned on simultaneously and 
the geometric reconnection of the brane is intimately related with the condensation 
of a normalizable mode of the bifundamental complex scalar field. There are no branes, in 
particular, with vanishing condensate of the complex scalar mode whose dynamics is 
captured solely by the DBI action.
\item[(2)] In section \ref{main} we set up an effective tachyon-DBI action that captures 
efficiently the interacting dynamics of the complex scalar field, the transverse scalars and 
abelian gauge field and reproduces key features of the exact string theory results in the 
NS5-D$p$-$\overline{{\rm D}p}$ system, $e.g.$ the exact CFT dependence of the boundary
cosmological constant on the hairpin-brane turning point.
In this description the duality of the worldsheet boundary interactions of the previous item 
acquires an intuitive geometric interpretation. The non-local nature of the complex 
scalar field is incorporated naturally. 
In our treatment of the tachyon-DBI action the tachyon potential is a free function.
We find an asymptotic tachyon-paperclip solution that reproduces some of the exact 
string theory information and solves the equations of motion with a modified potential,
$i.e.$ a potential different from the $1/\cosh$ one that has been derived in flat space.
\item[(3)] We clarify the role of the DBI action in this description. The deviations from the 
DBI solution are supported near the turning point of the hairpin-brane. We determine 
analytically the behavior of the tachyon-DBI solution near the turning point.
\item[(4)] We study how one incorporates the non-normalizable mode of the bifundamental
complex scalar field in the asymptotics of the solution. In the Sakai-Sugimoto model this mode 
is responsible for giving bare mass to the quarks. An issue in previous discussions of the 
Sakai-Sugimoto model was how to incorporate this mode without violating the boundary 
condition that the \ddbar\ tachyon vanishes at the asymptotic infinity. We will see that our 
effective description gets around this problem in an interesting way.  
\end{itemize}

Our ultimate goal is to import these lessons in more general contexts where no a priori 
worldsheet control is available. We are particularly interested in the Sakai-Sugimoto 
model and its implications for the strong coupling dynamics of QCD. A preliminary 
discussion of this system appears in section \ref{ss}. A more thorough examination
of the abelian TDBI description of the Sakai-Sugimoto model is currently under investigation
\cite{companion}.

\section{An abelian effective field theory description of the \ddbar\ system in flat space}
\label{ddbar}

In this section we review the key points of a new effective description of the
\ddbar\ system in flat space that was proposed in \cite{Erkal:2009xq}. This
effective description will be the basis of our subsequent analysis of hairpin-branes 
in holographic backgrounds.

Consider the action (henceforth we set $\alpha'=1$) 
\beq
\label{ddbaraa}
\SS=-\int d^{p+1} \sigma \, e^{-\Phi}V(T) \sqrt{-\det (G_{ab}+2\pi F_{ab})}
\eeq
where $\sigma^a$ ($a=0,1,\ldots,p$) are woldvolume coordinates, 
\beq
\label{ddbarab}
G_{ab}=g_{\mu\nu}\d_a X^\mu \d_b X^\nu+\d_a T \d_b T
\eeq
is the induced metric, $F_{ab}$ is the field strength of an abelian gauge field and  
$g_{\mu\nu}(X)$, $\Phi(X)$ are the spacetime metric and dilaton respectively. 
Formally, this is the DBI action for a $p$-brane propagating in a `fictitious' 
(10+1)-dimensional spacetime with an extra spatial coordinate $T$ and metric
\beq
\label{ddbarac}
ds^2=g_{\mu\nu} dX^\mu dX^\nu +dT^2
~.
\eeq
From this point of view the potential $V(T)$ appears as a $T$-dependent contribution to 
the dilaton field.

Ref.\ \cite{Erkal:2009xq} observes that the action \eqref{ddbaraa}, with potential $V(T)$ given by 
\eqref{introaac}, provides a unified description of BPS and non-BPS branes in flat space. The 
non-BPS D$p$-brane is obtained by orienting the worldvolume perpendicular to $T$. Indeed, 
by choosing the static gauge $X^a=\sigma^a$ ($a=0,1,\ldots, p$) one recovers the TDBI action 
\eqref{introaa}. The fields of this description are the worldvolume gauge field, the physical space 
transverse scalars, and the tachyon field $T$. 

The BPS D$(p-1)$-brane is obtained by choosing a different orientation of the $(p+1)$-dimensional
worldvolume in the extended spacetime ---an orientation {\it parallel} to $T$. Adopting the static gauge 
$X^a=\sigma^a$ ($a=0,1,\ldots,p-1$), $T=\sigma^p$ we recover an action for the transverse scalars
and $U(1)$ gauge field. With a $T$-dependent transformation one can set the $T$-component
of the gauge field, $A_T$, to zero. This leaves a residual symmetry of $T$-independent gauge
transformations. The remaining fields are all functions of both $\sigma^a$ and $T$, but Ref.\ 
\cite{Erkal:2009xq} shows that the general $T$-dependent configuration is non-normalizable
and does not describe open string excitations. In fact, one can argue that such configurations
describe closed string excitations in accordance with the open string completeness proposal of
Sen \cite{Sen:2004nf}. The bottom line of this discussion is that one should consider 
$T$-independent profiles of the fields $X^\mu, A_a$. Then, the action \eqref{ddbaraa} becomes
\beq
\label{ddbarad}
\SS=-\left( \int dT \,V(T)\right) \int d^p \sigma\, e^{-\Phi} 
\sqrt{-\det (g_{\mu\nu} \d_a X^{\mu} \d_b X^\nu +2\pi F_{ab})}
~.
\eeq
In flat space, with the tachyon potential \eqref{introaac}, the BPS brane tension is
\beq
\label{ddbarae}
\tau^{\rm BPS}_{p-1}=\int dT\, V(T)=\sqrt{2}\pi \tau_p^{\rm non-BPS}
\eeq
and \eqref{ddbarad} reduces nicely to the DBI action for a BPS D$(p-1)$-brane.
It is possible to generalize this discussion to include worldvolume fermions and Wess-Zumino
couplings \cite{Erkal:2009xq}. Since neither of these extra features will be important for our 
purposes we will not consider them here explicitly.

We will assume that the action \eqref{ddbaraa} is a sensible starting point for 
the effective field theory description of BPS and non-BPS D-branes in general spacetime 
backgrounds with a background-dependent tachyon potential $V(T)$. In section \ref{main} 
we will verify this assumption in a class of curved backgrounds where an exact string theory 
description is available. 

So far we have seen how standard BPS or non-BPS D-branes in flat space are reproduced 
by trivial planar solutions of the action \eqref{ddbaraa}. It is interesting to explore the physical 
meaning of more general solutions with different orientations/shapes in the extended 
(10+1)-dimensional spacetime. 

An interesting inhomogeneous solution of the TDBI equations of motion (in flat space and 
tachyon potential \eqref{introaac}) is described by the equation
\beq
\label{ddbaraf}
\sinh(\alpha T)=A \cos(\alpha x)
~.
\eeq
This solution is a Euclidean version of the rolling tachyon solution \cite{Sen:2004nf}.
$x$ is one of the physical space coordinates and $\alpha$ the constant that appears in the 
tachyon potential \eqref{introaac}. $A$ is a free real constant, $i.e.$ a modulus of the solution.

\begin{figure}[t!]
\centering
\includegraphics[height=7cm]{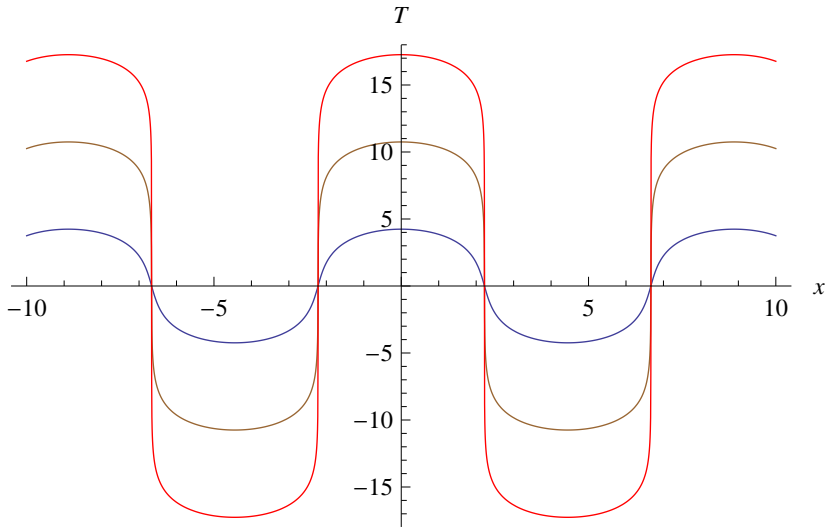}
\bf\caption{\it \small The euclidean rolling tachyon profile \eqref{ddbaraf} is plotted here in $(x,T)$
space for $\alpha=\frac{1}{\sqrt 2}$ and three different values of the free constant $A$. 
$A=10^2$ for the blue curve, $A=10^3$ for the brown curve and $A=10^5$ for the red curve.}
\label{rollprofile}
\end{figure}

For $A=0$ we recover a non-BPS brane oriented along $x$. For finite $A$ the open string 
tachyon of the non-BPS brane has condensed. On the worldsheet a marginal boundary
interaction of the form
\beq
\label{ddbarag}
\delta \LL_{ws}=\mu_B \cos(\alpha x)
\eeq
has been turned on.
As we increase the magnitude of $A$ the profile of the solution develops larger and larger
regions oriented along the $T$ direction (see Fig.\ \ref{rollprofile}). Hence, for $A=\pm \infty$ we 
recover an array of BPS-antiBPS branes separated by a distance
\beq
\label{ddbarai}
L_*=\frac{\pi}{\alpha}
\eeq
controlled by the TDBI parameter $\alpha$. For the type II string theory value 
$\alpha=\frac{1}{\sqrt 2}$ the separation is $L_*=\pi \sqrt 2$. This particular value of $L$ has  
a special significance from the point of view of the \ddbar\ system. When the D and 
$\overline {\rm D}$-branes are separated by this distance the complex \ddbar\ tachyon 
$\TT$ is massless. This fact is expressed naturally in the TDBI solution by the marginality 
of the parameter $A$. The TDBI action \eqref{introaa} recovers in this way a critical 
property of the open string theory of the \ddbar\ system.

Reversing the order of this observation it is also instructive to make the following point.
Assume that we knew that the TDBI action \eqref{introaaa} is a good description of open 
string dynamics for non-BPS branes, but we had no prior knowledge of the precise form of the 
tachyon potential $V(T)$. A quick way to determine the potential would have been to demand 
that the action 
\beq
\label{ddbaraj}
\SS=-\int dx\, V(T)\sqrt{1+\left(\frac{dT}{dx}\right)^2}
\eeq
has a one-parameter family of periodic solutions with a modulus-{\it independent} period 
(as in Fig.\ \ref{rollprofile}). One can easily verify that the only potential with this property is the 
$1/\cosh$ potential.

\begin{figure}[t!]
\centering
\includegraphics[height=8cm]{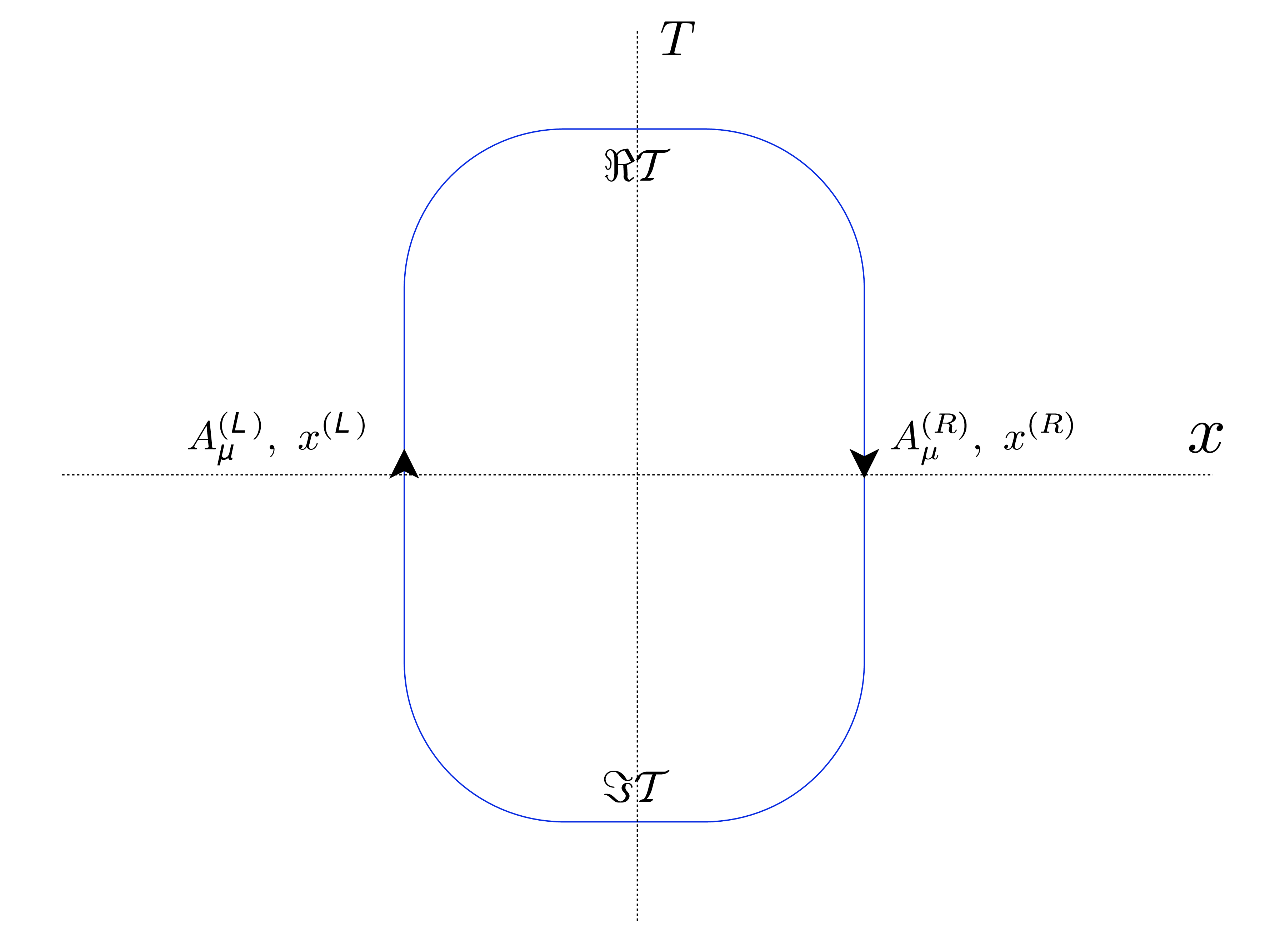}
\bf\caption{\it \small The tachyon paperclip profile. Different parts of its geometry capture different
degrees of freedom of the \ddbar\ system.}
\label{Tpaperclip}
\end{figure}

The above discussion demonstrates that the abelian TDBI action has the necessary ingredients 
to provide an effective description not only of the BPS and non-BPS branes, but also of the \ddbar\ 
system. For the latter it is more appropriate to consider a {\it closed} curve in $(x,T)$ space with a 
`paperclip' profile of the form depicted in Fig.\ \ref{Tpaperclip}. To first approximation we may 
view this profile as arising from the joining of a large positive-$A$ half-period of the solution 
\eqref{ddbaraf} (upper part) with a large negative-$A$ half-period of the same solution (lower part). 
When the $T$-width of the paperclip profile is large, the legs ---perpendicular to the $x$ direction--- 
represent a D-brane separated by a distance $L_*$ from a $\overline{\rm D}$-brane.
The low-energy degrees of freedom on each of these legs are an abelian gauge field, $A_a^{(L)}$
or $A_a^{(R)}$, and the corresponding transverse scalars, $x^{(L)}$ or $x^{(R)}$. The upper
and lower parts of the paperclip, associated with the $A$ mode of the TDBI solution, capture
the real and imaginary parts of the complex \ddbar\ tachyon $\TT$. Ref.\ \cite{Erkal:2009xq}
demonstrates explicitly how one recovers the full set of \ddbar\ modes from the single abelian 
gauge field $A_a$ and real scalar $T$ of the abelian TDBI description. 

A crucial part of this dictionary is a non-trivial transformation relating the 
\ddbar\ $\TT$ and the TDBI $T$. For $T\gg 1$ and $\TT$ real this transformation takes the form
\beq
\label{ddbarak}
\TT\sim A^{-1} \sim e^{-\alpha T}
\eeq
where $T$ is evaluated at its maximum on the cental vertical axis of the paperclip. The inversive 
nature of this relation expresses the anticipated feature that $\TT$ is mildly condensed when the 
maximum value of $T$ is large. Moreover, treating the paperclip amplitude $A$ as a slowly varying 
field one recovers an action of the form
\beq
\label{ddbaraka}
\SS=-\frac{\tau_{p-1}^{\rm BPS}}{2}
\int d^p x \d_\mu \phi \d^\mu \phi \sum_{n=0}^\infty a_n 
\left( \frac{\d_\mu \phi \d^\mu \phi}{\phi^2}\right)^n
\eeq
with calculable constants $a_n$ and $\phi\sim A^{-1/2}$ an appropriately normalized field. 
This expansion suggests that the TDBI description is better suited to large values of $\TT\sim \phi^2$. 

\begin{figure}[t!]
\centering
\includegraphics[height=8cm]{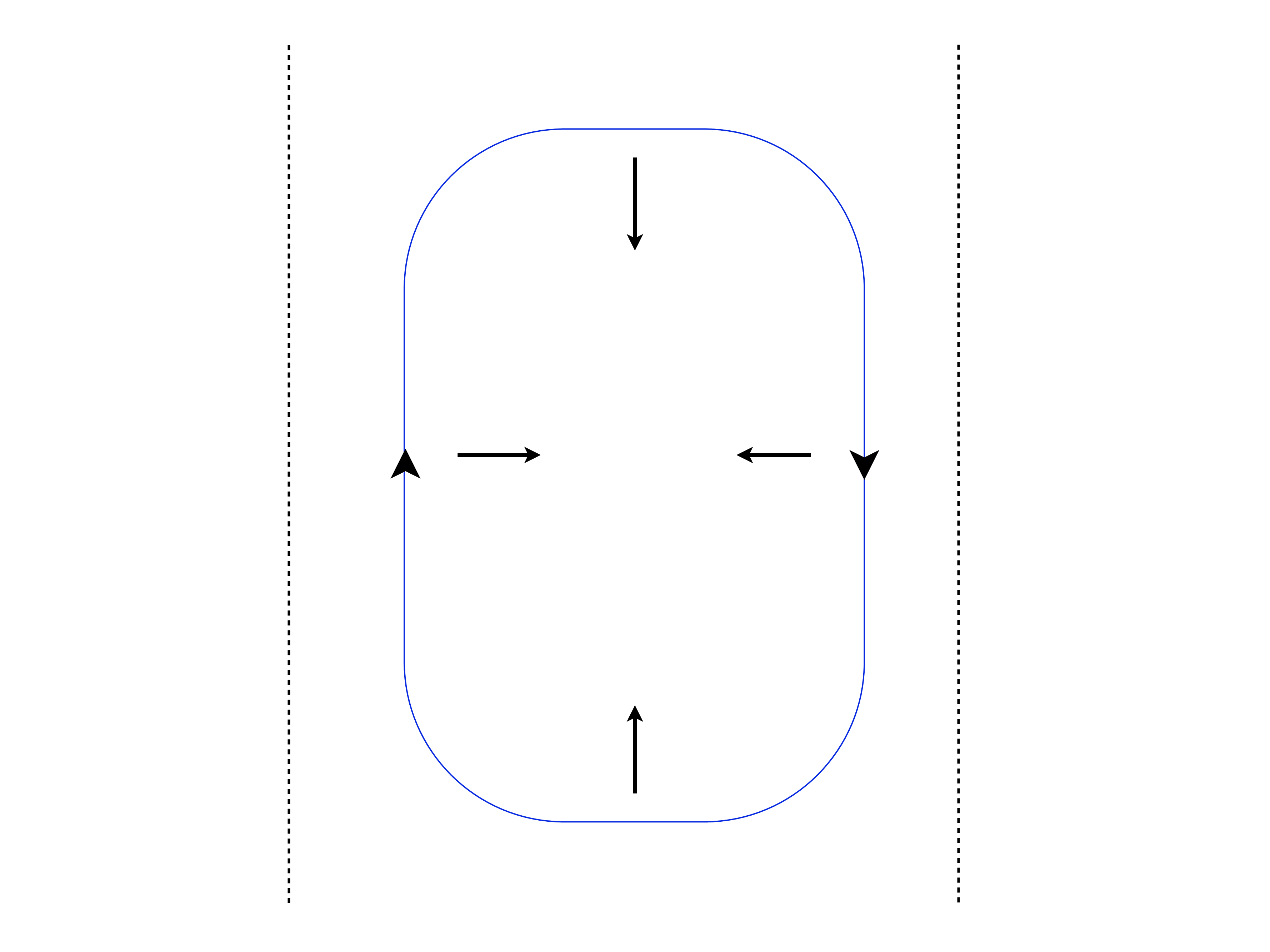}
\bf\caption{\it \small Condensation of the \ddbar\ tachyon in the TDBI language. For $L<L_*$ the 
tachyon-paperclip acquires a time-dependent evolution during which it shinks down to zero 
size.}
\label{Tcondensation}
\end{figure}

In this language, the real time process of \ddbar\ tachyon condensation takes a simple and geometric 
form. When the separation $L$ of the paperclip legs is larger than the critical separation $L_*$ 
(see eq.\ \eqref{ddbarai}), it is energetically favored for the paperclip to grow indefinitely in the 
$T$-direction leaving behind a \ddbar\ system with $\TT=0$. This property captures the fact that 
the \ddbar\ tachyon is massive in this case. In the opposite regime, $L<L_*$, it is energetically 
favored for the paperclip to shrink. There is a time-dependent solution that describes the 
tachyon-paperclip shrinking down to zero size (see Fig.\ \ref{Tcondensation}). In this process 
the legs of the paperclip move closer and closer. In this manner, the time-dependent 
real tachyon $T$ captures efficiently the complicated coupled dynamics of the complex \ddbar\ 
tachyon $\TT$ {\it and} the transverse scalars $x^{(L)}, x^{(R)}$. The non-local nature of $\TT$ 
is incorporated naturally in this description.

In what follows we will argue that there is a similar effective field theory description of the
open string dynamics on hairpin-branes in holographic backgrounds. Instead of a 
time-dependent tachyon-paperclip, in this case we have to consider a tachyon-paperclip
`condensing' along the radial direction of the holographic background.

\section{Hairpin-branes in holographic backgrounds}
\label{ssNS5}

We will concentrate on two specific examples of holography: the Sakai-Sugimoto 
model for QCD and a closely related example of NS5-branes that describes a 
non-supersymmetric version of the six-dimensional Little String Theory compactified 
on a circle. In this section we review the main features of these models (emphasizing
common properties and differences) and set up our notation.

\subsection{Lightning review of the Sakai-Sugimoto model}
\label{ssreview}

The Sakai-Sugimoto model \cite{Sakai:2004cn,Sakai:2005yt} is based on the following 
configuration of D4 and D8-branes in type IIA string theory
\beq
\label{SSaa}
\begin{array}{cccccccccccc}
      &  & 0 & 1 & 2 & 3 & 4 & 5 & 6 & 7 & 8 & 9\\
D4 &  :& \times & \times & \times & \times & \bullet &    &     &     &    &  \\
D8 &  :& \times & \times & \times & \times &      & \times & \times & \times & \times &\times \\
{\overline{D8}} &  :& \times & \times & \times & \times &      & \times & \times & \times & \times &\times
\end{array}
\eeq
The worldvolume theory on the $N_c$ D4-branes gives rise to a maximally supersymmetric
$U(N_c)$ gauge theory in five dimensions. The $x^4$ direction is compactified (hence the
solid circle in the D4 line of \eqref{SSaa}) with supersymmetry breaking antiperiodic
boundary conditions for the fermions. In the deep infrared (IR), below the Kaluza-Klein (KK) scale
set by the radius $R_4$, the dynamics is dominated by the four-dimensional Yang-Mills 
theory with gauge group $SU(N_c)$. The intersection of the D4-branes with $N_f$ D8-branes
supports chiral fermions in the fundamental representation of the gauge group. These fermions
propagate in the 3+1 dimensions common to both sets of branes. Similarly, the intersection with 
$N_f$ $\overline {{\rm D} 8}$-branes provides an analogous set of anti-chiral fermions. The D8 
and $\overline{ {\rm D} 8}$ stacks are separated by a distance $L$ in the $x^4$ direction. 

At weak 't Hooft coupling, the IR dynamics of this system is captured by a non-local version of the 
Nambu-Jona-Lasinio model \cite{Nambu:1961tp,Antonyan:2006vw}. In this paper we are more 
interested in the strong 't Hooft coupling regime. In the probe approximation, where $N_f/N_c\ll 1$,
this regime is captured holographically by $N_f$ D8-branes in the supergravity background 
of the Wick-rotated black D4-brane
\begin{subequations}
\label{SSabfull}
\beq
\label{SSab}
ds^2=\left(\frac{u}{R}\right)^{\frac{3}{2}}
\left( -dt^2+(dx^i)^2+f(u)(dx^4)^2 \right)
+\left(\frac{R}{u}\right)^{\frac{3}{2}}\left(\frac{du^2}{f(u)}+u^2 d\Omega^2_4\right)
~,
\eeq
\vspace{-0.3cm}
\beq
\label{SSac}
e^\Phi=g_s \left(\frac{u}{R}\right)^{\frac{3}{4}}~, ~ ~ ~ F_4=3\pi N_c \ell_s^3 \varepsilon_4
~,
\eeq
\beq
\label{SSad}
f(u)=1-\frac{u_{\rm KK}^3}{u^3}
~.
\eeq
\end{subequations}
This solution is completely fixed by three numbers: $\frac{u_{\rm KK}}{\ell_s}$, $\frac{R}{\ell_s}$
and $N_c$. The absence of a conical singularity at $u=u_{\rm KK}$ fixes the radius $R_4$ of 
$x^4$ to the value
\beq
\label{SSae}
R_4=\frac{2}{3} \frac{R^{\frac{3}{2}}}{u_{\rm KK}^{\frac{1}{2}}}
~.
\eeq 
$g_s$ is related to $\frac{R}{\ell_s}$ and $N_c$ through the relation
\beq
\label{SSaf}
g_s=\frac{1}{\pi N_c} \left(\frac{R}{\ell_s}\right)^3
~.
\eeq

At strong 't Hooft coupling curvatures are small everywhere, but the string coupling $e^\Phi$ 
becomes large in the asymptotic region $u\gg 1$. For that reason, one usually sets a strong 
coupling cutoff at 
\beq
\label{SSag}
u_{\rm max}\sim g_s^{-4/3} R
~.
\eeq

The D8 and anti-D8 branes reconnect in this background and form $N_f$ hairpin-like branes 
in the $(u,x^4)$ part of the geometry which has a cigar-like topology. Using the DBI action one finds 
a one-parameter family of solutions parameterized by the turning point value $u_0$. At 
$u\gg u_0$ the two branches of the D8-brane are separated by a distance $L$ which is a 
function of $u_0$. As we increase $u_0$, $L(u_0)$ decreases. When $u_0=u_{\rm KK}$, $i.e.$
when the brane reaches the tip of the $(u,x^4)$ cigar, the asymptotic separation $L$ becomes
maximal and the D8-brane is situated at antipodal points of the $x^4$ circle.

The geometric reconnection of the D8-branes is a nice feature of the DBI solution that expresses 
naturally the flavor chiral symmetry breaking that takes place in the dual gauge theory. 
Nevertheless, it is clear already on the basis of the general holographic dictionary that the 
DBI description cannot be the full story. In this description the order parameter of chiral symmetry 
breaking is absent and with it an important set of open string degrees of freedom that affect the
low-energy physics. For that reason a more appropriate description of the open string dynamics 
of the D8-branes is needed. We will return to propose such a description in section \ref{ss}.

\subsection{A Little String Theory analog of the Sakai-Sugimoto model}
\label{lst}

Before addressing the dynamics of hairpin-branes in the Sakai-Sugimoto model it is instructive
to consider a closely related situation where instead of $N_c$ D4-branes we have $k$
NS5-branes. Since there are only NSNS fluxes in the background geometry of this system an 
explicit worldsheet description of both closed and open string dynamics is possible with
the use of the RNS formalism. With this technical advantage we can use this system to 
set the standard lore: identify the crucial properties of string theory and test any proposals 
for an effective field theory that aspires to reproduce them.

Specifically, consider the following brane configuration in type IIB string theory
\beq
\label{lstaa}
\begin{array}{cccccccccccc}
      &  & 0 & 1 & 2 & 3 & 4 & 5 & 6 & 7 & 8 & 9\\
NS5 &  :& \times & \times & \times & \times & \bullet & \times &     &     &    &  \\
D1 &  :& \times &  &  &   &      &  & + &   &  &  \\
{\overline{D1}} &  :& \times &  &  &  &  &  & + &  &  &
\end{array}
\eeq
The worldvolume theory on $k\geq 2$ coincident NS5-branes gives rise at low energies to 
a six-dimensional Little String Theory \cite{Aharony:1999ks}, a very interesting interacting 
non-gravitational theory whose dynamics remains largely unknown. Once again we 
compactify the $x^4$ direction (one of the worldvolume directions of the NS5-branes) 
on a circle with radius $R_4$ imposing supersymmetry breaking antiperiodic boundary 
conditions for the fermions. A set of D1 and anti-D1 branes extend along the upper half of 
the $x^6$ direction (hence the $+$ in the second and third lines of \eqref{lstaa}) and 
intersect the NS5-branes at $x^6=0$. The D1-branes are separated from the anti-D1-branes 
by a distance $L$ along $x^4$. T-dualizing along the $x^1,x^2,x^3$ or $x^5$ directions 
we can trivially obtain D$p$ and anti-D$p$ branes intersecting the NS5-branes in type IIA 
or type IIB string theory. In order to be concrete we will restrict our discussion 
to D1-branes oriented as in \eqref{lstaa}.

Before compactification the configuration of the NS5-branes is supersymmetric
and the near-horizon geometry of $k$ NS5-branes is given by the CHS background 
\cite{Callan:1991at}
\begin{subequations}
\label{lstab}
\beq
\label{lstaba}
ds^2=\eta_{\mu\nu}dx^\mu dx^\nu+k(d\rho^2+d\Omega_3^2)
~,
\eeq
\vspace{-0.5cm}
\beq
\label{lstabb}
\Phi=-\rho~, ~~ H_3=dB_2 =k \, \varepsilon_3
\eeq
\end{subequations}
where $\varepsilon_3$ is the volume form of the transverse three-sphere. 
String theory on this background is described by an exact CFT on the worldsheet. String
propagation in the four-plane $(6789)$ transverse to the NS5-brane worldvolume  
is described by the $\NN=(1,1)$ supersymmetric linear dilaton CFT and the $SU(2)_k$ WZW 
model. The core of this throat, at large negative $\rho$, has a strong coupling singularity.

In this background the D1-branes are aligned along the linear dilaton direction $\rho$
and are pointlike in the transverse sphere. The separation $L$ of the D1 and anti-D1-branes 
in the $x^4$ direction is arbitrary. In the NS$-$ sector of the string stretching 
between a D1 and an anti-D1-brane there is a mode with mass squared 
\cite{Elitzur:2000pq,Israel:2005fn}
\beq
\label{lstac}
{\bf M}^2_\TT=\frac{L^2}{4\pi^2}-\frac{k}{4}
~.
\eeq
This mode becomes massless when the separation $L$ takes the critical value
\beq
\label{lstad}
L_*=\pi \sqrt{k}
~.
\eeq
We observe that the mass formula \eqref{lstac} receives a negative contribution 
proportional to $k$, which is potentially large when $k$ is large. The origin of this
contribution lies at the non-trivial dependence of the wavefunction along the linear 
dilaton direction. As a result, this particular mode can be massless even for separations
$L_*$ much larger than the string scale, when $k\gg 1$. The mode we are currently describing
is special for the following reason. For $L>L_*$ there are no tachyonic modes
in the open string theory of the D1-${\overline{\rm D1}}$ system. As we lower $L$ below $L_*$ the 
above mode is the first one to become tachyonic.

After the compactification with supersymmetry breaking boundary conditions, the background
of the NS5-branes becomes that of the Wick-rotated black NS5-brane. In the near-horizon limit
\begin{subequations}
\label{lstae}
\beq
\label{lstaea}
ds^2=-dt^2+\sum_{i=1,i\neq 4}^5 (dx^i)^2+k d\Omega_3^2
+k \left( d\rho^2+\tanh^2\rho \, d\theta^2\right)
\eeq
\vspace{-0.5cm}
\beq
\label{lstaeb}
e^\Phi=\frac{g_s}{\cosh\rho}~, ~~ H_3 =k\, \varepsilon_3
~.
\eeq
\end{subequations}
$\rho$ is now a positive real number and $\theta=\frac{x^4}{\sqrt{k}}$. 
To avoid a singularity at $\rho=0$, $\theta$ needs to be compactified, $i.e.$ $\theta\sim \theta+2\pi$.
With $g_s\ll 1$, string theory on this background is everywhere perturbative and one 
can apply perturbative string theory techniques to study its properties. String propagation
on the cigar geometry parametrized by $(\rho,\theta)$ is described on the worldsheet by
an exact CFT, the $\NN=(2,2)$ $SL(2)_k/U(1)$ gauged WZW model \cite{Witten:1991yr}. 
Curvatures are small compared to the string scale when $k\gg 1$, but since we have an 
explicit string theory description of this system we can discuss its properties for any $k\geq 2$.

As in the case of the Sakai-Sugimoto model, the D1-${\overline{\rm D1}}$ pairs reconnect
to form hairpin D1-branes that extend along the radial direction of the two-dimensional 
cigar and the $U(1)\times U(1)$ symmetry, which is local on the D1-branes, is broken to a 
diagonal $U(1)$. There is a corresponding two-parameter family of hairpin-brane solutions 
of the DBI action described by the embedding equation \cite{Fotopoulos:2003vc}
\beq
\label{lstaf}
\theta(\rho)=\theta_0+\arcsin \left(\frac{\sinh\rho_0}{\sinh\rho}\right)
~.
\eeq
$\rho_0$ is the turning point of the brane and $\theta_0$ is a trivial shift 
of the asymptotic angular position of the hairpin branches. The asymptotic separation $L$ 
is $\rho_0$-independent and takes automatically the critical value $L_*$ that we encountered in 
eq.\ \eqref{lstad}. According to the mass formula \eqref{lstac},\footnote{This formula remains valid 
in the non-supersymmetric compactified background, see for instance \cite{Antoniadis:1998ki}.} 
there is a half-winding mode of the \ddbar\ tachyon which is massless. This bifundamental mode, 
whose vev breaks the $U(1)\times U(1)$ symmetry, plays an important role in the open string 
dynamics of these branes as we proceed to review in the next subsection.
 
An interesting and consequential difference with the Sakai-Sugimoto case, which is worth 
highlighting here, can be traced to the DBI fact of modulus-dependence (in the Sakai-Sugimoto
model) or modulus-independence (in the NS5 model) of the asymptotic separation $L$. Apart from 
this detail, gross features of hairpin-brane open string dynamics are expected to be common
in these systems.

\subsection{Open string theory in the near-horizon region of NS5-branes}
\label{cft}

The D1 hairpin-branes that we have just described have an exact treatment in string theory
as boundary states in the $SL(2)/U(1)$ CFT 
\cite{Ribault:2003ss, Eguchi:2003ik,Ahn:2003tt,Israel:2004jt,Ahn:2004qb,Fotopoulos:2004ut}. 
Here we summarize their most pertinent properties.

The axial coset $SL(2)_k/U(1)$ that describes string propagation on the cigar-shaped
background
\beq
\label{cftaa}
ds^2=k \left( d\rho^2+\tanh^2\rho\, d\theta^2 \right)~, ~~
\Phi=\Phi_0-\log\cosh\rho
\eeq
is a Kazama-Suzuki model \cite{Kazama:1988qp} with $\NN=(2,2)$ worldsheet supersymmetry 
and central charge $c=3+\frac{6}{k}$. By mirror symmetry it is equivalent to the $\NN=2$ Liouville 
theory \cite{Giveon:1999px,Hori:2001ax}. The spectrum of closed and open string modes is 
organized according to the representation theory of $SL(2)/U(1)$. The worldsheet primary fields 
are labeled by five quantum numbers: $j$, $m$, $\bar m$, $s$, $\bar s$ (bars denote right-moving 
sector quantities). We will denote the corresponding vertex operators as
\beq
\label{cftab}
\VV^{(s,\bar s)}_{j,m,\bar m} ~~~({\rm or~} \VV^{(s)}_{j,m}~ {\rm for~the~chiral~version})
~.
\eeq
The triplet $(j,m,\bar m)$ is a triplet of $SL(2)$ quantum numbers. $j$ is the $SL(2)$
spin and $m,\bar m$ are directly related to the momentum and winding quantum numbers
($n,w$ respectively) in the angular direction $\theta$ of the geometry
\beq
\label{cftac}
m=\frac{kw+n}{2}~, ~~ \bar m=\frac{kw-n}{2}
~.
\eeq
In the mirror $\NN=2$ Liouville theory the roles of $n$ and $w$ are exchanged. The $(s,\bar s)$ 
are worldsheet fermion numbers ($s=0$ for the NS sector and $s=1$ for the R sector). In $\NN=2$
Liouville language the chiral version of the above vertex operators reads
\beq
\label{cftad}
\VV^{(s)}_{j,m}=e^{-j\rho+i(m+s)\theta+isH}
\eeq
where $H$ is a worldsheet boson that bosonizes a complex fermion. The scaling dimension
of this operator is
\beq
\label{cftae}
\Delta_{j,m}^{(s)}=\frac{(m+s)^2-j(j-1)}{k}+\frac{s^2}{2}
~.
\eeq
Observe the possibly negative contribution of the Casimir $-j(j-1)$ to the scaling dimension
(and hence the mass squared of the corresponding spacetime mode) for real values of the spin.

The allowed values of the spin $j$ depend on the representation of $SL(2)$. In the closed
string spectrum two representations appear: {\it continuous} representations with
\beq
\label{cftaf}
j=\frac{1}{2}+ip~, ~~ p\in \R
\eeq
and {\it discrete representations} with
\beq
\label{cftag}
\frac{1}{2}<j<\frac{k+2}{2}~, ~~ \pm m=j+\Z_{>0}
~.
\eeq
The continuous representations provide delta-function normalizable modes. $p$, the imaginary
part of $j$, can be viewed as momentum in the radial direction $\rho$ of the background. The 
discrete representations provide normalizable modes supported near the tip of the cigar 
(these modes can be viewed as localized twisted-sector modes, to use a language familiar 
from orbifolds). What we want to capture eventually with an effective field theory is the dynamics 
of a mode of the \ddbar\ tachyon that comes from such a discrete representation.

There is a feature of closed string theory in this background that we want to emphasize before 
proceeding to review the open string theories of interest. The background metric and dilaton 
\eqref{cftaa}, which solve the leading order (in $\alpha'$) supergravity equations, do not receive
any $\alpha'$ corrections in perturbation theory and are exact to all orders in $1/k$
\cite{Bars:1992sr,Tseytlin:1993my}. From the point of view of the asymptotic cylinder, at 
$\rho\to \infty$, the cigar-shaped `bending' of the geometry appears as a worldsheet interaction 
with vanishing momentum and winding of the form
\beq
\label{cftai}
\tilde \mu_{bulk} \int d^2 z\, (\d H-k\d\theta)(\bar \d H-k \bar \d \theta)e^{-\rho}
~.
\eeq
This interaction appears as a D-term deformation in the $\NN=(2,2)$ worldsheet Lagrangian.

It is well known, however, that string theory on this background receives non-perturbative
corrections in the form of a `winding tachyon condensate'. This condensate, which is none  
other than the winding $\NN=2$ Liouville interaction, is an F-term deformation of the form
\beq
\label{cftaj}
\mu_{bulk}\int d^2 z \, e^{-\frac{k}{2}(\rho+i\tilde \theta)+i(H+\bar H)}+{\rm c.c.}
~.
\eeq
It involves the vertex operators $\VV^{(\pm 1,\pm 1)}_{\frac{k}{2},\mp \frac{k+2}{2},\mp \frac{k+2}{2}}$.
$\tilde \theta$ is the $T$-dual of the angular coordinate $\theta$.
One can easily verify, using the general formula \eqref{cftae}, that these vertex operators have
scaling dimensions (1,1) and give rise to marginal interactions on the worldsheet, $i.e.$ massless
fields in spacetime. It should be appreciated that the strengths of these interactions 
$(\tilde \mu_{bulk}; \mu_{bulk},\bar \mu_{bulk})$ are not independent parameters and that these 
interactions appear simultaneously in the worldsheet CFT.\footnote{The precise relation between 
these parameters is not important here. It can be found for example in \cite{Hosomichi:2004ph}.} 
From the worldsheet point of view these interactions play the role of dual screening charges
\cite{Ahn:2002sx} (see also \cite{Nakayama:2004vk}). Their simultaneous presence is intimately 
related to the mirror symmetry equivalence between the $\NN=2$ cigar coset and the $\NN=2$ 
Liouville theory. Mirror symmetry converts the non-perturbative winding $\NN=2$ Liouville
condensate of the cigar coset to the standard momentum tachyon condensate of the
$\NN=2$ Liouville theory. We will now see that a similar picture of dual worldsheet 
interactions characterizes the D1 hairpin-branes of interest.

In the worldsheet boundary CFT the D1 hairpin-branes can be formulated as boundary 
states $|J,M\rangle$ labeled by two parameters \cite{Israel:2004jt,Hosomichi:2004ph}. 
$M$ is a parameter related to $\theta_0$ in \eqref{lstaf} and has trivial physics. Here we 
will set it to zero. $J$ takes the complex value $\frac{1}{2}+iP$. $P$, 
which is a non-negative real number, is related to the turning point $\rho_0$ of the DBI profile 
\eqref{lstaf} (see eq.\ \eqref{numacaa} below for a more precise version
of this relation). In particular, $P=0$ when $\rho_0=0$.

The parameters $(J,M)$ control the strength of the boundary worldsheet interactions
that characterize the open string theory on these branes. The details of these interactions in 
$\NN=2$ Liouville language can be found in Ref.\ \cite{Hosomichi:2004ph} (see section
5.1 in that paper). Two kinds of boundary worldsheet interactions are simultaneously turned on. 
The first one is a holomorphic square root of the bulk interaction \eqref{cftai} proportional to
\beq
\label{cftak}
\tilde \mu (\d H-k\d\theta)e^{-\rho}
~.
\eeq
The second one is a holomorphic square root of the bulk interaction \eqref{cftaj} proportional to
\beq
\label{cftal}
\mu e^{-\frac{k}{2}(\rho +i\tilde \theta)+iH}
\eeq
together with its complex conjugate. In terms of the brane label $P$
\beq
\label{cftam}
\mu \bar \mu \sim \sinh(2\pi P)~, ~~ \tilde \mu \sim \sinh \left( \frac{2\pi P}{k} \right)
~.
\eeq
Boundary fermions are needed to express accurately these interactions (see 
Ref.\ \cite{Hosomichi:2004ph} and appendix \ref{cosmo}). Note that the first relation in \eqref{cftam}
is only valid for non-vanishing $P$. In fact, $\mu$ and $\bar \mu$ are non-zero for $P=0$.
More details can be found in appendix \ref{cosmo}.

These interactions can be viewed as open string analogs of \eqref{cftai} and \eqref{cftaj}.
The first interaction \eqref{cftak} has zero winding and captures the geometric 
bending of the D1-brane. As we will see, this is the piece of the open string theory captured 
quite well in the asymptotic region by the DBI action. The second interaction is
non-geometric and involves a massless half-winding mode of the \ddbar\ tachyon.
This is the mode we need to incorporate in the effective field theory description of these
branes extending the DBI action. The present explicit string theory example clarifies how 
this bifundamental mode appears in the open string spectrum and why it is relevant for 
open string dynamics. In particular, notice that none of the hairpin-branes has a vanishing 
condensate of this mode and none of these branes can be described fully by the DBI action. 

To summarize the lessons of the string theory analysis,
we conclude that the chiral symmetry breaking $U(1)_L\times U(1)_R\to U(1)_{diag}$ 
that one observes on the D1-branes geometrically as a reconnection effect is not unrelated 
to the physics of the \ddbar\ tachyon which is the right field to carry the order parameter for 
this breaking. String theory knows at a fundamental level about both of these aspects and 
incorporates them simultaneously. In order to capture them efficiently with an effective field 
theory description one has to write an action that generalizes the DBI action and incorporates 
the effects of the \ddbar\ tachyon. We will see that this is naturally achieved by the abelian TDBI 
action outlined in section \ref{ddbar}. We are going to claim that this picture is not specific to the 
example of NS5-branes and that a qualitatively similar picture applies also to other situations, 
$e.g.$ the Sakai-Sugimoto model.

\section{Effective field theory description of D-branes intersecting NS5-branes}
\label{main}

This is the main section of this paper. Our primary task is to set up an effective field
theory description that reproduces the exact open string theory features that were
summarized above.

\subsection{The TDBI action and its equations of motion}
\label{tdbi}

In order to set up an effective action for the D1 hairpin-brane in the background of NS5-branes
we will start, following the idea of \cite{Erkal:2009xq}, from the tachyon-DBI action of a non-BPS 
D2-brane that wraps the two-dimensional cigar geometry \eqref{cftaa}.\footnote{The branes of 
interest are point-like in the remaining transverse directions, $e.g.$ the transverse 3-sphere,
so we can ignore them.} Applied to this background the TDBI action 
\eqref{ddbaraa} reads (with vanishing gauge field)
\beq
\label{tdbiaa}
\SS=-\int d\rho d\theta \, \sinh \rho\, V(T) \sqrt{1+\frac{1}{k}(\d_\rho T)^2
+\frac{1}{k}\coth^2\rho (\d_\theta T)^2}
~.
\eeq
Since we want to consider profiles of the tachyon that are not single-valued functions of 
$\theta$ it will be also convenient to choose a different parametrization
\beq
\label{tdbiab}
T=T(\rho,\sigma)~, ~~ \theta=\theta(\rho,\sigma)~, ~~ \sigma\in [0,2\pi)
~.
\eeq
In this parametrization the TDBI action becomes
\beq
\label{tdbiac}
\SS=-\int d\rho d\sigma \, \sinh\rho\, V(T) \,
\sqrt{(\d_\sigma \theta)^2+\frac{\coth^2\rho}{k}(\d_\sigma T)^2
+\frac{1}{k}\left(\d_\rho \theta \d_\sigma T-\d_\rho T \d_\sigma \theta\right)^2}
~.
\eeq

The Euler-Lagrange equations of the action \eqref{tdbiaa} can be written in the form
\beq
\label{tdbiad}
\d_\rho \left( \frac{1}{k} \sinh\rho \frac{V}{\sqrt W} \d_\rho T \d_\theta T\right)
=\d_\theta \left( \sinh\rho \frac{V}{\sqrt W} \left( 1+\frac{1}{k}(\d_\rho T)^2 \right) \right)
\eeq
where 
\beq
\label{tdbiae}
W\equiv 1+\frac{1}{k}(\d_\rho T)^2+\frac{1}{k}\coth^2\rho (\d_\theta T)^2
~.
\eeq
Varying the action \eqref{tdbiac} with respect to $T$ and $\theta$ we obtain one
independent differential equation of the form
\beq
\label{tdbiaf}
\d_\rho \left( \sinh\rho \frac{V}{\sqrt{Q}}
\left(\d_\rho \theta \d_\sigma T-\d_\rho T \d_\sigma \theta\right)\d_\sigma T \right)
=\d_\sigma \left( \sinh\rho \frac{V}{\sqrt{Q}} \left(-k\d_\sigma \theta+
\left(\d_\rho \theta \d_\sigma T-\d_\rho T \d_\sigma \theta
\right) \d_\rho T\right)\right)
\eeq
where
\beq
\label{tdbiag}
Q\equiv (\d_\sigma \theta)^2+\frac{\coth^2\rho}{k}(\d_\sigma T)^2
+\frac{1}{k}\left(\d_\rho \theta \d_\sigma T-\d_\rho T \d_\sigma \theta\right)^2
~.
\eeq

We are looking for solutions of the equations of motion of these actions with non-trivial profiles of 
$T=T(\rho,\theta)$ (or $T=T(\rho,\sigma)$, $\theta=\theta(\rho,\sigma)$)
that behave at large $\rho$ like the tachyon-paperclip profile of Fig.\ \ref{Tpaperclip}.
Instead of condensing in time, the tachyon-paperclip will now condense in the radial
direction of the holographic background. The elementary degrees of freedom of the D1 
hairpin-brane, $e.g.$ the transverse scalar and \ddbar\ tachyon associated to the worldsheet 
boundary interactions \eqref{cftak} and \eqref{cftal} respectively, will emerge in this description 
along the lines of the discussion in section \ref{ddbar}.

In \eqref{tdbiaa} we left the tachyon potential $V(T)$ as an undetermined function.
In flat space, a derivation from first principles \cite{Kutasov:2003er,Niarchos:2004rw} fixed the 
potential to the $1/\cosh$ form \eqref{introaac}. Motivated by the universality of the form of the DBI 
action, it would seem natural to propose that there is no background dependence in the tachyon 
potential and that the $1/\cosh$ form should continue to apply in general curved backgrounds. 
Nevertheless, both in the derivation of \cite{Kutasov:2003er,Niarchos:2004rw} and in the discussion 
of boundary string field theory \cite{Witten:1992qy,Kutasov:2000qp} the tachyon potential is closely 
related to the disc partition function of string theory which is a quantity that depends explicitly on the 
closed string background over which we are computing. Furthermore, in the computation of 
\cite{Kutasov:2003er,Niarchos:2004rw} the disc partition function is calculated in the background of 
the rolling tachyon solution, whose specifics are also expected to be dependent on the closed string 
background. Since it is currently unclear how the background will affect the tachyon potential in these 
computations, we will adopt a strategy in which we treat $V(T)$ as a free function. Our goal is to find
tachyon-paperclip solutions of the general TDBI equations that reproduce some of the key 
features of the exact string theory answer. In what follows we will achieve this goal with solutions 
that require the modification of the tachyon potential for generic $k$. At the same time, we observe 
that the TDBI equations with the $1/\cosh$ potential \eqref{introaac} admit (in a specific regime) 
a solution with features that are not expected from string theory for $k>2$. These results could be 
viewed as an indication that the $1/\cosh$ potential is indeed modified in curved backgrounds, 
however, one would need a more comprehensive analytical control of the TDBI equations to make 
a conclusive argument on this issue. We do not claim to have a complete argument of this sort in 
this paper.

\subsection{Constraining the tachyon potential: $|T|\ll 1$}
\label{Tsmall}

\begin{figure}[t!]
\centering
\includegraphics[height=6cm]{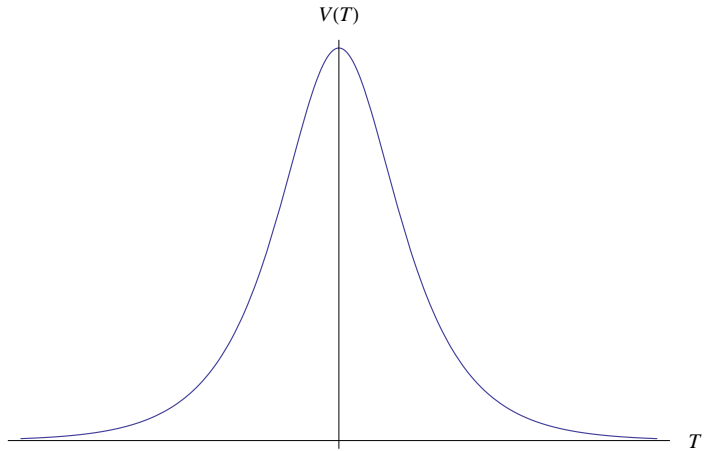}
\bf\caption{\it \small A graphical depiction of the general form of the tachyon potential $V(T)$.}
\label{V(T)}
\end{figure}

With the above considerations in mind, we assume that the tachyon potential is an analytic function 
of the general form depicted in Fig.\ \ref{V(T)}. It is symmetric under $T\to -T$, it has a maximum at 
$T=0$, and it decreases monotonically with increasing $|T|$ towards zero for $|T|\to \infty$. A Taylor 
expansion of the potential around $T=0$ gives to quadratic order
\beq
\label{Tsmallaa}
V(T)\simeq V(0)\left(1-\frac{1}{2}\alpha^2 T^2+\OO(T^4) \right)
\eeq
where we have set
\beq
\label{Tsmallab}
\frac{1}{V(0)}\frac{d^2V}{dT^2}\Big|_{T=0}=-\alpha^2
~.
\eeq
$V(0)$ expresses the tension of the non-BPS brane (here a non-BPS brane wrapping
the cigar). $\alpha^2$ is directly related to the mass ${\bf M}_T$ of the real tachyon that resides on 
the non-BPS brane.

To obtain the precise relation between $\alpha^2$ and ${\bf M}_T$ we expand the TDBI action
\eqref{tdbiaa} up to quadratic order in $T$ and $\d T$
\beq
\label{Tsmallac}
\SS\simeq -V(0)\int d\rho d\theta\, \sinh\rho 
\left( 1+\frac{1}{2k}(\d_\rho T)^2+\frac{\coth^2\rho}{2k} (\d_\theta T)^2-\frac{1}{2}\alpha^2 T^2 \right)
+\ldots
~.
\eeq
Defining a renormalized tachyon
\beq
\label{Tsmallad}
S=\sqrt{\cosh \rho}\, T
\eeq
the equation of motion of the quadratic action \eqref{Tsmallac} becomes
\beq
\label{Tsmallae}
\Box S- {\bf M}_T^2 S=0
\eeq
with
\beq
\label{Tsmallaf}
{\bf M}_T^2=-\alpha^2+\frac{1}{k}\left( 1-\frac{3}{4}\tanh^2\rho\right)
~.
\eeq 
At the asymptotic infinity $(\rho\to \infty)$ we read off the tachyon mass squared
\beq
\label{Tsmallag}
{\bf M}_T^2(\infty)=-\alpha^2 +\frac{1}{4k}
~.
\eeq

In the open string theory of the non-BPS D2-brane the NS$-$ sector mode of interest 
has vanishing angular momentum and winding $(n=w=0)$ and belongs to the continuous
representation with radial momentum $p=0$, $i.e.$ $SL(2)$ spin $j=\frac{1}{2}$. The mass
of this tachyonic mode can be read off the formula
\beq
\label{Tsmallai}
{\bf M}_T^2=\Delta^{(0)}_{\frac{1}{2},0}-\frac{1}{2}=-\frac{1}{2}+\frac{1}{4k}
\eeq
where eq.\ \eqref{cftae} was used to determine the second equality.
Matching this expression to the TDBI result \eqref{Tsmallag} requires
\beq
\label{Tsmallaj}
\alpha=\frac{1}{\sqrt{2}}
~,
\eeq
a $k$-independent value that matches the one encountered for non-BPS branes in type II string 
theory in flat space. Notice that the mass formula \eqref{Tsmallai} reproduces correctly the 
flat space result as we send $k\to \infty$ and make the cigar geometry arbitrarily weakly curved.

\subsection{Constraining the tachyon potential: $|T|\gg 1$ and paperclip asymptotics}
\label{asymptotics}

In the previous subsection we have determined the leading order quadratic behavior
of the tachyon potential around the open string vacuum. We will now discuss the 
behavior of the potential around the closed string vacuum at $|T|\gg 1$. Our main 
purpose is to find a sensible tachyon-paperclip solution of the TDBI equations at 
$\rho \to \infty$ and to match the $\rho$-dependence of this solution to the exact 
information \eqref{cftal} from open string theory. We will achieve this purpose with 
the assumption of a specific ansatz for the large-$T$ behavior of the tachyon-paperclip 
solution. We will discover that this ansatz requires a modification of the tachyon potential
\eqref{introaac}. The possibility of keeping the potential unchanged will be
discussed briefly at the end of this subsection.

A motivated guess for the large-$T$ behavior of the tachyon potential is
\beq
\label{asyaa}
V(T)\sim e^{-\beta T}~, ~~ T\gg 1
~.
\eeq
For the potential \eqref{introaac} $\beta=\alpha=\frac{1}{\sqrt 2}$. In what follows, however, we will 
keep $\beta$ as a free positive number that will be adjusted appropriately to satisfy the equations
of motion. The negative $T$ behavior of $V$ is controlled by the symmetry $V(T)=V(-T)$. In order to 
match the exact information \eqref{cftal} from string theory we look for a large-$T$ paperclip solution 
of the TDBI action with large-$\rho$ asymptotics of the form
\beq
\label{asyab}
T\sim a+b\rho+c\log\cos\theta
~.
\eeq
As a simple modification of the large-$T$ behavior of the Wick-rotated version of the rolling tachyon 
solution in flat space (see section \ref{ddbar}) this ansatz has automatically the qualitative features that 
we are after. The $\rho$-dependence has been chosen in a way that will soon allow us to reproduce
the exact information \eqref{cftal} from string theory.
We have also fixed the would-be legs of the tachyon-paperclip solution to lie at 
diametrically opposite points of the cigar $S^1$, $\theta=\pm \frac{\pi}{2}$, as required from string 
theory. A trivial constant shift, which can be added freely to $\theta$, will not be written out 
explicitly here. In this profile we want $a$ to be an arbitrary constant that signals the presence
of a marginal half-winding mode. The parameters $b$ and $c$ are $k$-dependent constants that 
need to be determined. 

Inserting the ansatz \eqref{asyab} into the TDBI equations of motion \eqref{tdbiad}
we obtain
\beq
\label{asyac}
\left[ \beta-\frac{b}{k} \left(\beta b-1\right)\left(1+\frac{b^2}{k}\right)^{-1} \right]
\left[ \left(1+\frac{b^2}{k}\right)\cos^2\theta +\frac{c^2}{k}\sin^2\theta\right]=\frac{c}{k}
~.
\eeq
The constant $a$ drops out as expected and remains free.
This equation can be satisfied for general $\theta$ if and only if 
\beq
\label{asyad}
c^2=b^2+k
\eeq
in which case the $\theta$-dependence disappears and we end up with the extra equation 
\beq
\label{asyae}
\beta=\frac{1}{b+c}
~.
\eeq
Once we fix from string theory the $\rho$-dependence of $T$, namely the constant $b$, 
we can determine the remaining constants, $c$ and $\beta$, from eqs.\ \eqref{asyad} and 
\eqref{asyae}.

We can read off the asymptotic $\rho$-dependence of the \ddbar\ tachyon from the 
vertex operator $\VV$ \eqref{cftal}. Defining an appropriately normalized version of 
this vertex operator, $\TT$, we obtain
\beq
\label{asyaf}
\TT \equiv e^{d \rho}\VV \sim \mu\, e^{-(\tilde b-d)\rho}+\ldots+M\, e^{(\tilde B+d)\rho}+\ldots 
\eeq
with
\beq
\label{asyag}
\tilde b=\frac{k}{2}~, ~~ \tilde B=\frac{k}{2}-1
~.
\eeq
We are keeping explicit only the $\rho$-dependent piece of the vertex operator here
and include both the normalizable branch with proportionality coefficient $\mu$ and the 
non-normalizable branch with coefficient $M$. Note that in the corresponding expansion
of the \ddbar\ tachyon in the Sakai-Sugimoto model, $\mu$ would express holographically 
the order parameter of chiral symmetry breaking and $M$ would be related to the bare 
quark mass. The constant $d$ that appears in the normalization factor $e^{d\rho}$ will be 
fixed appropriately in a moment.

In the asymptotic region, where $T\gg 1$, the exponential of $T$ has a corresponding
$\rho$-dependent expansion at $\theta=0$
\beq
\label{asyai}
e^T\sim \mu_T e^{b\rho}+\ldots+M_T e^{-B\rho}+\ldots~, ~~ b\geq 0, ~~ b+B>0
~.
\eeq
The first (leading) term in this expansion is captured already by eq.\ \eqref{asyab}.
The exponent of the first term of the subleading series, with proportionality constant $M_T$,
can be determined from the TDBI equation of motion \eqref{tdbiad} using a general-$k$
version of the differential equation \eqref{numab} below (specifically, eq.\ \eqref{enumab}
in appendix \ref{extranumerics}). In terms of the leading exponent $b$ we find
\beq
\label{asyaj}
B=1-b+\frac{b^2}{k}
~.
\eeq

The next task is to determine the relation between the two expansions \eqref{asyaf}
and \eqref{asyai}. Using the results of appendix \ref{tachyons} we deduce that
there is a non-trivial inversive transformation between $T$ and $\TT$ of the form 
\eqref{ddbarak}
\beq
\label{asyak}
\TT\sim e^{-\gamma T} ~~ \Leftrightarrow ~~ e^T\sim \TT^{-\frac{1}{\gamma}}
~.
\eeq
$\gamma$ is a positive number related to the $U(1)_L-U(1)_R$ charge of $e^T$.
We will soon see how $\gamma$ is constrained.

At this point it is useful to recall the following mathematical identity
\beq
\label{asyal}
\frac{1}{\pi} \left( \frac{1}{|z|^2 e^\phi+e^{-\phi}}\right)^{2h}=
\frac{1}{2h-1}e^{2(h-1)\phi}\delta^2(z)+\OO(e^{2(h-2)\phi})+
\frac{e^{-2h\phi}}{\pi |z|^{4h}}+\OO(e^{-2(h+1)\phi})
~.
\eeq
The rhs of this equation is a large-$\phi$ expansion. This expansion 
appears, for instance, in the discussion of primary fields with spin $h-1$ in 
the Euclidean $SL(2,\C)/SU(2)$ WZW model \cite{Kutasov:1999xu}. The 
subleading term $e^{-\phi}$ in the denominator of the lhs can play an important role 
when $|z|^2$ is sufficiently small. This effect gives rise to the first series in the rhs of 
\eqref{asyal}.

Applying the identity \eqref{asyal} to the leading order terms of $\TT^{-\frac{1}{\gamma}}$
we obtain
\bea
\label{asyam}
&&e^T\sim \TT^{-\frac{1}{\gamma}}\sim 
\left( \frac{1}{Me^{(\tilde B+d)\rho}+\ldots+\mu e^{-(\tilde b-d) \rho}+\ldots} \right)^{\frac{1}{\gamma}}
\nonumber\\
&&\simeq \frac{\pi \gamma}{1-\gamma} \mu^{1-\frac{1}{\gamma}}
e^{\left( (\frac{1}{\gamma}-1)(\tilde b-d)-(\tilde B+d)\right)\rho}\delta(M)+\ldots+ 
M^{-\frac{1}{\gamma}} e^{-\frac{\tilde B+d}{\gamma}\rho}+\ldots
~.
\eea
Matching the leading terms of both branches in the expansions \eqref{asyam}
and \eqref{asyai} we obtain the following relations
\begin{subequations}
\label{asyan}
\beq
\label{asyana}
b= \left(\frac{1}{\gamma}-1\right)(\tilde b-d)-(\tilde B+d)
~,
\eeq
\beq
\label{asyanb}
B=\frac{\tilde B+d}{\gamma}
~.
\eeq
\end{subequations}
Given $\tilde b$, $\tilde B$ (see eq.\ \eqref{asyag}) and $\gamma$ we can solve these 
equations (together with \eqref{asyaj}) to obtain the unknown parameters $b,c,d$ and $\beta$.
In terms of $\gamma$ we find
\begin{subequations}
\label{asyao}
\beq
\label{asyaoa}
b=\sqrt{\frac{k}{\gamma}} \sqrt{k-1-k\gamma}~, ~~
\beta=\sqrt{\frac{\gamma}{k}}\frac{1}{\sqrt{k-1-k\gamma}+\sqrt{(k-1)(1-\gamma)}}
~,
\eeq
\beq
\label{asyaob}
c^2=k(k-1)\frac{1-\gamma}{\gamma}~,~~
d=\frac{1}{2}\left(k+2(1-k)\gamma-\sqrt{k\gamma}\sqrt{k-1-k\gamma}
\right)
~.
\eeq
\end{subequations}

The above expressions for $b$, $\beta$ and $d$ are real and finite if and only if
\beq
\label{asyap}
0<\gamma \leq 1-\frac{1}{k}
~.
\eeq
The upper bound of this inequality is saturated when $T(\rho,0)$ \eqref{asyab} asymptotes to 
a constant. In that case,
\beq
\label{asyar}
b=0~, ~~ c=\sqrt{k}~, ~~ d=2-\frac{1}{k}-\frac{k}{2}~, ~~
B=1~, ~~ \beta=\frac{1}{\sqrt{k}} ~~{\rm and}~~ \gamma=1-\frac{1}{k} 
~.
\eeq
For these values and $k=2$ the function $\cosh^{-1}(\frac{T}{\sqrt 2})$ poses as a 
good candidate for the full tachyon potential $V(T)$. Indeed, this potential would 
reproduce correctly in this case both the small-$T$ and large-$T$ asymptotics, 
\eqref{Tsmallaa}, \eqref{Tsmallaj} and \eqref{asyaa}, \eqref{asyar} respectively. 
We recall that this is also the tachyon potential for non-BPS branes in flat space in 
type II string theory \eqref{introaac}. In favor of this identification we note that 
similarities between the system of two NS5-branes and string theory in flat space 
have been observed before \cite{Kutasov:2004ct}. For generic $k>2$, the $1/\cosh$
potential cannot reproduce both the small-$T$ and large-$T$ asymptotics that were
postulated above.

It is tempting to speculate that the expressions \eqref{asyar} give the correct tachyon 
asymptotics for all values of $k$. Additional information is needed, however, to derive 
this result analytically. Strong numerical evidence in favor of this expectation is provided
in appendix \ref{extranumerics}.

One may also wonder whether it is possible to reproduce the available exact string theory information
with more complicated tachyon-paperclip solutions and other tachyon potentials.
For instance, we did not analyze the possibility of a more complicated tachyon-paperclip
solution that does not require the modification of the $1/\cosh$ potential \eqref{introaac}.
Irrespective of the existence of such solutions, notice that in the particular case of \eqref{introaac} 
the flat space solution \eqref{ddbaraf} continues to solve the TDBI equations of motion to leading 
order at $\rho \to \infty$. For $\alpha=\frac{1}{\sqrt 2}$ and $k>2$ this fact 
seems to imply that there is a tachyon-paperclip solution that represents the asymptotic \ddbar\ 
branches of a hairpin-brane with the wrong separation $L<\pi \sqrt k$. If this is true and the 
above solution can be properly extended to a regular $\rho$-dependent tachyon-paperclip 
solution beyond the large-$\rho$ region it would appear to invalidate the action \eqref{tdbiaa} 
with potential \eqref{introaac} as a sensible description of D1 hairpin-branes for general $k>2$; 
it would be a solution that represents a set of D1-branes that do not exist in the background of 
$k$ NS5-branes.

\subsection{Normalizable vs non-normalizable modes of the \ddbar\ tachyon}
\label{norm}

The non-trivial transformation \eqref{asyam} has another interesting consequence.
Comparing the expansions \eqref{asyai} and \eqref{asyam} we also learn that
\beq
\label{normaa}
\mu_T\sim \mu^{1-\frac{1}{\gamma}}\delta(M)~, ~~
M_T\sim M^{-\frac{1}{\gamma}}
~.
\eeq
The roles of leading versus subleading branches are exchanged under the 
transformation \eqref{asyak}. The coefficient $\mu$ that controls the subleading 
series in the expansion of the \ddbar\ tachyon \eqref{asyaf} controls in terms of $T$ 
the coefficient of the leading series and vice versa for the coefficient $M$ that 
controls the leading series in the expansion of the \ddbar\ tachyon.
In the language of holographic QCD, the branch associated to chiral 
symmetry breaking in the dual gauge theory becomes leading in the 
TDBI formulation. Adding a non-zero quark mass leads to a modification
of the subleading branch that does not destroy the natural asymptotic boundary
condition $T\to \infty$ (equivalently $\TT\to 0$) at $\rho\to \infty$. 

In previous discussions of tachyon dynamics in the context of the Sakai-Sugimoto 
model \cite{Bergman:2007pm,Dhar:2008um} one encounters the issue of how to 
compromise this boundary condition with the presence of the leading branch that 
appears when we give bare mass to the quarks. The exponential divergence of the 
non-normalizable branch of the \ddbar\ tachyon is simply incompatible with a vanishing 
$\TT$ at infinity and one has to resort into a delicate limiting prescription of how to make 
sense of a solution that incorporates holographically the bare quark mass \cite{Dhar:2008um}. 

Here we learn that this is a problem related more to the description rather than the fundamentals 
of the D-brane system. In the TDBI description both branches can be incorporated 
simultaneously and without any conflict with the natural asymptotic boundary conditions. 
Of course, if we insist to make the identification \eqref{asyam} strictly at infinity we 
rediscover the issue in the form of a delta function contribution to the coefficient of the leading 
order branch \eqref{normaa}. In holographic applications what matters is the precise 
dictionary between bulk and boundary quantities. It would be interesting to elaborate further
on a more direct dictionary between TDBI variables and dual gauge theory quantities.

\subsection{The solution near the turning point}
\label{turning}

So far we have discussed the asymptotic, $\rho\to \infty$, behavior of the paperclip solution, which 
is controlled to a large degree ($i.e.$ sufficiently away from the paperclip legs) by the
large-$T$ properties of the tachyon potential. Now we want to discuss the behavior of the
paperclip near the turning point of the hairpin-brane. This behavior is controlled by the small-$T$ 
properties of the tachyon potential.

To capture the tachyon-paperclip behavior in this region we will use the parametrization 
\eqref{tdbiab} and solve the partial differential equation \eqref{tdbiaf}, \eqref{tdbiag} perturbatively 
around the turning point $\rho_0$. We are looking for a capping-off solution with the properties
\begin{subequations}
\label{turnaa}
\bea
&&|T| \ll 1~, 
\\
&&|\d_\rho T|, |\d_\rho \theta|\gg 1~,
\\
&&|\d_\sigma T|, |\d_\sigma \theta|\ll 1~,
\\
&&0<\rho-\rho_0\ll 1
~.
\eea
\end{subequations} 
Expanding in powers of $\rho-\rho_0$ we set
\beq
\label{turnab}
T(\rho,\sigma)= \sqrt{\rho-\rho_0} \Big(T_0(\sigma)+ (\rho-\rho_0) T_1(\sigma)+\ldots\Big)~, ~~
\theta(\rho,\sigma)= \sqrt {\rho-\rho_0} \Big(\theta_0(\sigma)+ (\rho-\rho_0) \theta_1(\sigma)+\ldots\Big)
~.
\eeq
We will see in a moment that the leading $\sqrt{\rho-\rho_0}$ behavior facilitates a natural 
solution for generic $\rho_0\neq 0$. We insert this ansatz into the differential equation \eqref{tdbiaf} 
and keep all terms up to the next-to-leading order in $\rho-\rho_0$. The expansion of the tachyon 
potential reads
\beq
\label{turnac}
V(T)=V(0)+\frac{1}{2}V''(0) T^2+\OO(T^4)=1-\frac{1}{2}  \alpha^2 (\rho-\rho_0) T_0^2
+\OO\left((\rho-\rho_0)^2\right)
~.
\eeq

The leading zero-th order term of the differential equation \eqref{tdbiaf} is automatically satisfied.
At the next-to-leading order we find
\beq
\label{turnad}
2\coth\rho_0 \, P_0\, \d_\sigma T_0 
-k\,\d_\sigma T_0 \frac{R_0}{P_0}
+2k \,\d_\sigma^2\theta_0
+\frac{k}{2} T_0 \frac{\d_\sigma R_0}{P_0}
=0
\eeq
where
\beq
\label{turnae}
P_0\equiv \frac{1}{2}\Big( \theta_0\d_\sigma T_0-T_0\d_\sigma \theta_0 \Big)
~,
\eeq
\beq
\label{turnaf}
R_0\equiv (\d_\sigma \theta_0)^2+\frac{\coth^2\rho_0}{k}(\d_\sigma T_0)^2
~.
\eeq
There are two interesting things happening here. First, the second order expansion 
of $T$ and $\theta$ does not appear in the second order expansion of the equation of motion.
Second, only the zero-th order (tension) term of the tachyon potential contributes.
The tachyon mass term that depends on $\alpha$ drops out. Hence, the leading order 
behavior of the solution, expressed in terms of the functions $T_0$ and $\theta_0$, is 
independent of the details of the tachyon potential and acquires a more robust and universal 
character.

In order to solve the equation \eqref{turnad} we make the following ansatz
\beq
\label{turnag}
T=A\sqrt{\rho-\rho_0} \cos\sigma+\ldots~, ~~
\theta=B\sqrt{\rho-\rho_0} \sin\sigma+\ldots
~.
\eeq
With this ansatz the $\sigma$-dependence drops out of the equation and what remains provides 
a polynomial relation between the constants $A$, $B$ and $\tanh \rho_0$
\beq
\label{turnai}
2k B^2 \tanh^2\rho_0-A^2 B^2 \tanh\rho_0+2A^2=0
~.
\eeq
Hence, there are two parameters, say $A$ and $B$, that control the solution. They
map at the asymptotic infinity to the two independent coefficients $\mu_T$ and $M_T$ 
(or $\mu$ and $M$) of the tachyon behavior. For a given pair of parameters $(A,B)$
there are two turning points solving the quadratic equation \eqref{turnai}
\beq
\label{turnaj}
\tanh \rho_0=\frac{A}{4kB}\left( AB\pm \sqrt{A^2 B^2-16 k} \right)
~.
\eeq
The positivity of the discriminant requires $A^2B^2\geq 16 k$. In general, when the 
discriminant is strictly positive, the two solutions are distinct and related formally by a simple 
parity transformation $AB\to -AB$ (or equivalently $\sigma\to -\sigma$). There is a single
solution when the discriminant vanishes, $i.e.$ when $A^2B^2=16k$. The existence of 
two independent solutions for the turning point implies that the hairpin-brane is parametrized
by two continuous parameters, $A$ and $B$ or $\mu$ and $M$, and an additional 
$\Z_2$ parameter. Since there is no known extra discrete parameter for vanishing $M$
we conjecture that the standard hairpin-branes reviewed in section \ref{cft} are described
by the vanishing discriminant case with
\beq
\label{turnak}
\tanh \rho_0=\frac{A^2}{4k}~, ~~ B^2=\frac{16k}{A^2}
~.
\eeq

Another interesting feature of the relation \eqref{turnai} is the breakdown of the expansion
\eqref{turnab}, \eqref{turnag} as we take the limit $\rho_0\to 0$. In that case, $A$ is forced 
towards zero and $B$ towards infinity making the leading order elliptical shape \eqref{turnag}
highly asymmetrical along the $\theta$-axis. This feature is reminiscent of the discontinuity
that we observe in the boundary cosmological constants $\mu,\bar \mu$ in worldsheet 
CFT as we take $P\to 0$ (see appendix \ref{cosmo}).

\subsection{Numerical results}
\label{numeric}

Given the precise form of the tachyon potential one would like to solve the full equations 
\eqref{tdbiaf}, \eqref{tdbiag} and determine the entire profile of the condensing tachyon-paperclip 
solution. Since an analytic solution appears to be out of reach it would be desirable to explore a 
numerical evaluation of the solution. The optimal strategy would be to shoot from the region near 
the turning point, using the universal ($i.e.$ tachyon potential independent) profile \eqref{turnag}, 
towards the asymptotic infinity. We will not pursue this exercise in this paper. Instead, we will 
explore numerically some of the properties of the solution using an approximate tool that simplifies 
the differential equation to an equation for a single-variable function. This equation will play a similar
role for the tachyon that the DBI equations play for the transverse scalars.

We will be setting boundary conditions at the asymptotic infinity. Estimates of the turning point will 
be provided by two independent sources: the singularity structure of the above approximate tool 
and the corresponding DBI solution. Already at this level of approximation we will recover a 
picture that corroborates the statements of the previous sections.

For illustration purposes we will fix the value of $k$ in the rest of this subsection to $k=2$
(see, however, appendix \ref{extranumerics} for a numerical analysis of the case with general
$k$). Although this is a highly stringy regime for the closed string background we anticipate that
the TDBI action will continue, even in this case, to provide a sensible description. The numerical 
results verify this expectation. Our main motivation for considering this value of $k$ is
that the tachyon potential $\cosh^{-1}(\frac{T}{\sqrt 2})$ poses as a good candidate
in this case. 

\begin{figure}[t!]
\centering
\includegraphics[height=5.2cm]{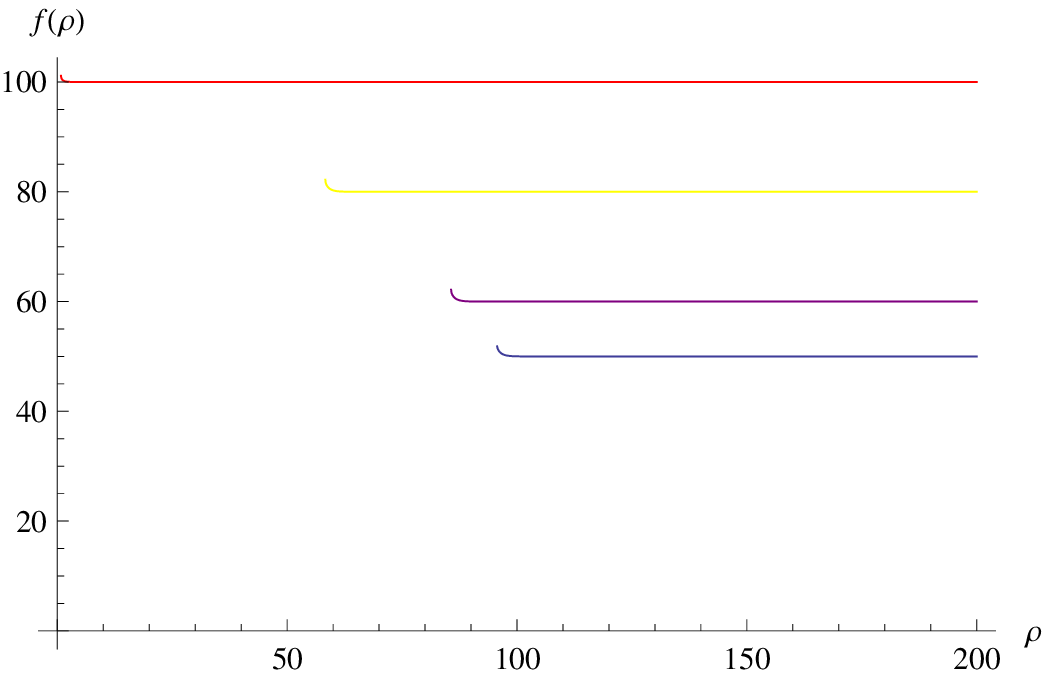}
\hfill
\includegraphics[height=5.2cm]{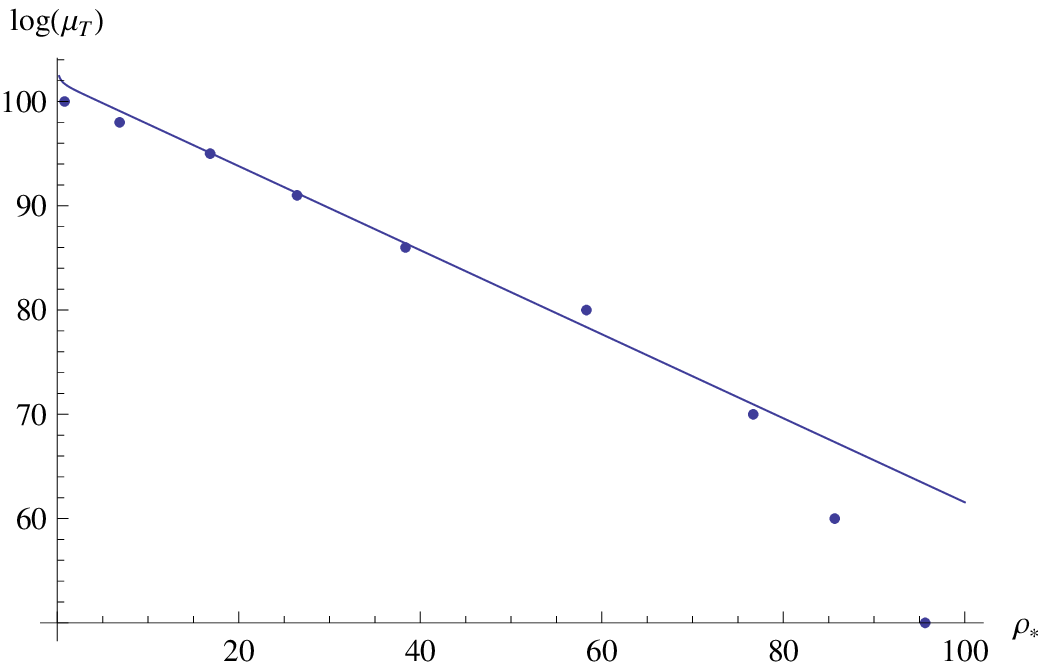}
\bf\caption{\it \small The left figure depicts the numerical evaluation of the differential equation 
\eqref{numab} with four different boundary conditions set at $\rho=100$. Each boundary condition is set
by the coefficients $\mu_T$ and $M_T$ in \eqref{asyai}. All curves have the same $M_T=e^{146}$,
a large value that implies via \eqref{normaa} $|M|\ll 1$. From the blue to the red curve we have 
$\mu_T=e^{50}, e^{60}, e^{80},e^{100}$. The curves terminate on the left at the point $\rho_*$ 
where the numerical evaluation encounters a singularity. The singularity occurs respectively at 
$\rho=95.6352,85.6667,58.3069,0.8018$ from the blue to the red curve. The nine plotted points 
in the right figure are based on a similar calculation with the same $M_T$ and express 
$\log(\mu_T)$ as a function of the singularity point $\rho_*$. The solid blue curve is a fit obeying 
the equation \eqref{numaca}, \eqref{numacaac}. In the vicinity of the tip of the cigar, $\rho_*\lesssim 1$,
both the solid blue curve and the numerically determined curve $\log(\mu_T)$ exhibit a vertical
rise to arbitrarily large values.}
\label{Avar}
\end{figure}

For starters, it is straightforward to check that the flat space solution \eqref{ddbaraf}, with 
$\alpha=\frac{1}{\sqrt 2}$, is an exact leading order solution of \eqref{tdbiad} at the asymptotic 
infinity. Since the profile of this solution is essentially that of a square function for large values 
of the amplitude (see Fig.\ \ref{rollprofile}) this profile, with a $\rho$-dependent $A$, is expected 
to give a sensible estimate of the tachyon maximum $T(\rho,0)$ as long as $T(\rho,0)$ is large 
enough and the legs of the tachyon-paperclip are not converging fast on each other. Accordingly, 
we make the following ansatz for the tachyon
\beq
\label{numaa}
T(\rho,\theta)=\sqrt{2} \, {\rm arcsinh} \left( \cos\theta \sinh \frac{f(\rho)}{\sqrt 2} \right)
~,
\eeq
we insert this ansatz into the equation of motion \eqref{tdbiad}, expand to leading order
around $\theta=0$ and obtain a differential equation for the function $f(\rho)$
\beq
\label{numab}
\left( f' \cosh\rho -\frac{\sqrt 2}{\sinh\rho} \tanh \left(\frac{f}{\sqrt 2}\right) \right)
\left( 2+{f'}^2 \right)+2\sinh\rho f''=0
~.
\eeq
We claim that this function provides a useful first estimate of $T(\rho,0)$ for a sufficiently large range 
of $\rho$.

We evaluated numerically the solution of the differential equation \eqref{numab} for different 
boundary conditions with the use of {\sc mathematica}. The result is plotted in Fig.\ \ref{Avar}.
Keeping the asymptotic coefficient $M_T$ (see eq.\ \eqref{asyai}) fixed and large we vary
the coefficient $\mu_T$. According to the transformation \eqref{asyam}, this corresponds
to keeping $M$ fixed and small and varying $\mu$. For each value of $\mu_T$ the 
numerical evaluation gives a function that remains essentially constant until it encounters 
a singularity at a point $\rho=\rho_*$ where the solution terminates. We will soon argue that 
$\rho_*$ provides a good estimate of the turning point of the brane ---an estimate that is 
comparable to the turning point predicted by the DBI solution. 

As we increase $\mu_T$, $i.e.$ as we go from the blue to the red curve in the left plot of 
Fig.\ \ref{Avar}, the predicted turning point is moving closer and closer to the tip of the cigar 
geometry. This is precisely the behavior we expect to see. As we increase $\mu_T$, $\mu$ 
decreases according to the first equation in \eqref{normaa}. Smaller $\mu$ implies a turning 
point closer to the tip of the cigar as we can deduce from the first CFT relation in \eqref{cftam}. 

In CFT the turning point $\rho_0$ can be determined in terms of the brane label $P$ in at least 
two different ways. One way is to define it implicitly by the mass of the ground state of an open 
string that stretches between 
a pointlike D0-brane at the tip of the cigar and the D1 hairpin-brane. Unfortunately, the spectrum 
of D0-D1 strings has not been computed explicitly in open string theory (such a computation 
involves the more complicated A-B type cylinder amplitudes, in the nomenclature of 
Ref.\ \cite{Fotopoulos:2004ut}, which have not been studied sufficiently). However, the spectrum 
of D0-D2 strings, for D2-branes characterized by a similar label $P$ as the D1-branes, has 
been studied extensively (see, for example, \cite{Fotopoulos:2004ut,Hosomichi:2004ph}). 
For generic $k$, the mass squared of the ground state of a D0-D2 string contains the term 
$\frac{P^2}{k}$. A similar term is expected for the ground state of the D0-D1 string. Assuming 
this is correct, equating this term with the flat space tension term $\frac{k \rho_0^2}{4\pi^2}$ 
gives the following relation between $P$ and $\rho_0$
\beq
\label{numacaa}
P=\frac{k \rho_0}{2\pi}
~.
\eeq
An alternative way to derive this relation, which produces an identical result, is based on a 
comparison of the semiclassical one-point functions of closed string vertex operators on the disc 
with the corresponding exact CFT one-point functions \cite{Ribault:2003ss,Fotopoulos:2004ut}.

\begin{figure}[t!]
\centering
\includegraphics[height=5.1cm]{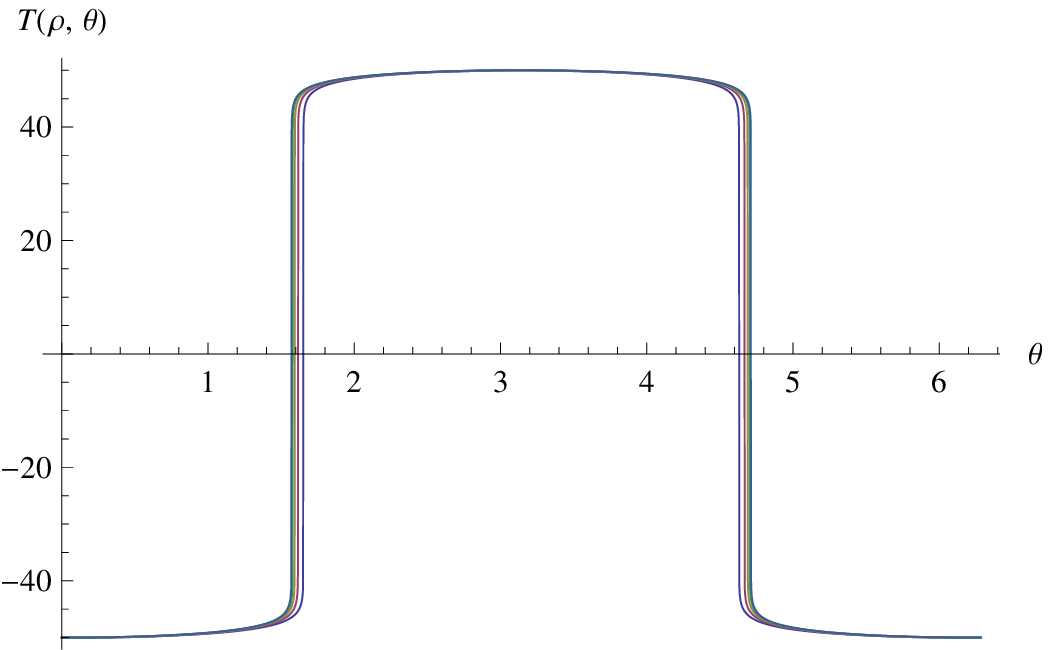}
\hfill
\includegraphics[height=5.1cm]{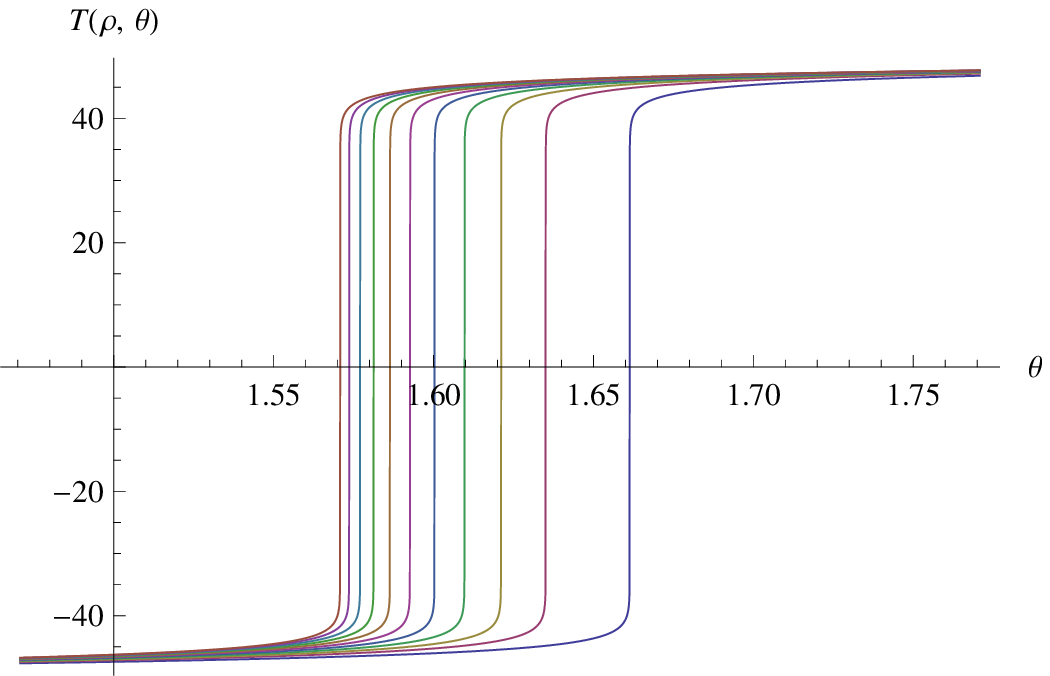}
\bf\caption{\it \small A plot of the tachyon profile $T(\rho,\theta)$ for several values of $\rho$
($\rho=97.9,98.2,98.4,\ldots,99.8,100$). Different values of $\rho$ are represented by 
different colors. In this particular example we are solving numerically the full equation 
of motion \eqref{tdbiad} using the profile \eqref{numaa} as a boundary condition at 
$\rho=100$ with radial derivative that gives $\mu_T=e^{50}$, $M_T=e^{146}$. The right 
figure provides a magnification of the solution near the position of one of the legs in the 
vicinity of $\theta=\pi/2$. We can see that with decreasing $\rho$ the legs move towards 
the center at $\theta=\pi$.}
\label{Trho}
\end{figure}

Identifying $\rho_*$ with the CFT $\rho_0$ and then combining the first equation in 
\eqref{normaa}, the first equation in \eqref{cftam} (with $\mu=\bar\mu$) and the 
relation \eqref{numacaa} we expect our numerical results to obey a relation of the form
\beq
\label{numaca}
\log \mu_T(\rho_*)=x-z \log\sinh(y \rho_*)
~,
\eeq
where $x,y,z$ are constants for which we anticipate analytically
\beq
\label{numacaab}
y=k=2~, ~~
z=\frac{1}{2}\left(\frac{1}{\gamma}-1\right)=\frac{1}{2(k-1)}=0.5
~.
\eeq
The precise value of $x$ depends on the proportionality constant that appears in the
first equation of \eqref{normaa}. This constant depends on the precise coefficients of the relation
\eqref{asyak} which we have left undetermined.

The right plot in Fig.\ \ref{Avar} exhibits nine numerically determined values of the function 
$\log\mu_T(\rho_*)$. Amusingly, the function \eqref{numaca} provides a rather good fit of 
these data with fit values\footnote{This fitting has been performed with the {\tt FindFit} 
command in {\sc mathematica} using five numerically determined data points at 
$\log(\mu_T)=70,80,86,91,95$.}
\beq
\label{numacaac}
x=101.507~, ~~ y=0.7941~, ~~ z=0.5075
~.
\eeq
The fit worsens as we approach the radius of the boundary conditions (in Fig.\ \ref{Avar} this 
is $\rho=100$). This is an expected deviation. The closer the turning point to the point of the 
boundary conditions the worse our approximation \eqref{numaa} becomes. In the opposite case,
$i.e.$ when the turning point is very close to the tip of the cigar, $\log(\mu_T)$ increases vertically 
in accordance with \eqref{numaca}. Strictly at $\rho_*=0$ (or $P=0$) the first relation in \eqref{cftam} 
(and therefore also \eqref{numaca}) breaks down (see appendix \ref{cosmo}).

\begin{figure}[t!]
\centering
\includegraphics[height=6cm]{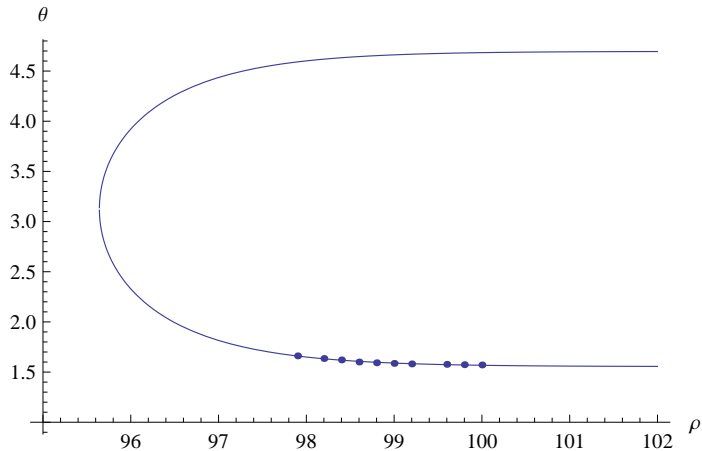}
\bf\caption{\it \small Fitting the motion of the tachyon-paperclip legs with the DBI solution 
\eqref{lstaf}.}
\label{DBIfit}
\end{figure}

The fitted value of $z$, 0.5075, compares well with the analytic value 0.5. There is a
greater mismatch for $y$. We expect $y=2$ from eq.\ \eqref{numacaa} and obtain 0.7941. 
At the same time, note that by focusing on $\rho_*$ values in the rough range from $\rho_*\sim 10$ to 
$\rho_* \sim 80$, as we did in obtaining the fit values \eqref{numacaac} (see footnote 8),
we can approximate \eqref{numaca} by the linear relation
\beq
\label{numacaad}
\log(\mu_T)=(x+z\log 2)-zy\, \rho_*
~.
\eeq
In this range we are essentially fitting to determine a single parameter, the slope $zy$. The value 
anticipated by \eqref{numacaab} is $zy=1$. The value produced numerically by the fit 
is $zy\simeq 0.4$. We observe a rough numerical agreement which was not guaranteed
to occur. We view this agreement, together with the qualitatively good behavior of the
TDBI-derived curve $\log(\mu_T)(\rho_*)$ in the vicinity of the tip of the cigar and away
from it, as non-trivial evidence for the consistency of our picture, the relevance of the 
approximate equation \eqref{numab} and the fact that $\rho_*$ provides a sensible 
estimate of the actual turning point of the brane. A similar picture arises at generic values
of $k$ (see appendix \ref{extranumerics} for further details).

Further evidence for the approximate identification of $\rho_*$ with the turning point of the brane is
provided by comparing $\rho_*$ to the turning point $\rho_{\rm DBI}$ predicted by the DBI 
solution. The bending described by the DBI solution is expected to deviate from the full 
TDBI result only within a small region around the turning point (see also the discussion in the 
following subsection \ref{dbi}). In that case, $\rho_{\rm DBI}$ is another, independent, estimate 
of the CFT turning point $\rho_0$ and should compare well with $\rho_*$. In Fig.\ \ref{Trho} we 
have plotted the profile of $T(\rho,\theta)$ solving the full equation of motion \eqref{tdbiad} with 
boundary conditions at $\rho=100$ corresponding to $\mu_T=e^{50}$, $M_T=e^{146}$.\footnote{The 
use of such a low value for $\mu_T$ is not ideal given the deviations of the fit in the right plot of 
Fig.\ \ref{Avar}. Our choice here is dictated purely by practical reasons related to the numeric
evaluation of the solution of the full equation of motion \eqref{tdbiad}.}
The $f$-estimate for the turning point of this brane was $\rho_*=95.6352$. In Fig.\ \ref{DBIfit} we plot 
ten numerically determined positions of the paperclip legs, $i.e.$ ten values of $\theta$ for which 
$T(\rho,\theta)=0$, as a function of $\rho$. We fit these values with the DBI solution \eqref{lstaf} to 
obtain 
\beq
\label{numad}
\rho_{\rm DBI}=95.6425
\eeq
a value that indeed compares well with $\rho_*=95.6352$.

These numerical results suggest that the `condensing' tachyon-paperclip solution has
a cigar-like shape in $(T,\rho,\theta)$ space that exhibits large radial gradients only near the
turning point region. As we change the asymptotic coefficients $\mu$, $M$, this shape
changes and capping-off occurs at different radial distances.

\subsection{Concluding comments on the role of the DBI action}
\label{dbi}

We conclude this section with a few general comments on the role of the DBI action.
We argued that the full low-energy dynamics of hairpin-branes in the background of 
NS5-branes cannot be described fully by the DBI action. We verified explicitly
in worldsheet conformal field theory that these branes always have a non-vanishing
condensate of a half-winding scalar mode that comes from the ground state of the NS$-$
sector. Irrespective of the asymptotic separation of the hairpin legs, which is controlled
by the number $k$ of NS5-branes, this mode is always massless. Its mass
squared receives large negative contributions from the non-trivial dependence of
the wavefunction in the radial direction of the background. A consistent treatment of
the dynamics of this field requires a different effective field theory treatment. We 
argued along the lines of Ref.\ \cite{Erkal:2009xq} that the abelian tachyon-DBI action 
for non-BPS branes provides an effective field theory description that captures correctly 
the main features of open string dynamics of a single hairpin-brane.

Nevertheless, we can still use the DBI action as a partial description of the system that 
captures sufficiently well the dynamics of the transverse scalars ($i.e.$ the geometric 
bending of the brane) in the asymptotic region where the legs of the tachyon-paperclip 
are long in the $T$-direction (see, for example, the fitting in Fig.\ \ref{DBIfit}.)
Deviations from the DBI result occur more prominently in a region near the turning
point where the tachyon-paperclip shrinks down to zero size. To determine these deviations for 
general $\mu$ and $M$ one needs to employ the full TDBI action. Moreover, we have 
seen that the tachyon-paperclip takes an elliptical shape near the turning point and 
that already at the classical level even the notion of a well-defined transverse space 
position for the brane is lost in this region. This is in accordance with the fact that the \ddbar\ 
tachyon is nearing there the closed string vacuum. The effective smearing of the brane near 
the turning point and the deviation from the DBI result are features that were also emphasized 
in Ref.\ \cite{Kutasov:2005rr}.

It is interesting to note that the turning point region is also a delicate region for the DBI 
action itself. The DBI solution \eqref{lstaf}
\beq
\label{dbiaa}
\theta(\rho)=\theta_0+\arcsin \left( \frac{\sinh\rho_0}{\sinh\rho} \right)
\eeq
develops large gradients near the turning point. For example, the second derivative
\beq
\label{dbiab}
\frac{d^2\theta}{d \rho^2}=\frac{1}{2}\sqrt{\frac{\coth(\rho_0)}{2}}(\rho-\rho_0)^{-\frac{3}{2}}
+\OO((\rho-\rho_0)^{-1/2})
\eeq 
diverges at $\rho \sim \rho_0$ and large gradients develop quickly in a region 
around the turning point. 

We conclude by highlighting some of the implications of the intimate relation between the 
dynamics of the transverse scalars and the \ddbar\ tachyon. This relation can be used to extract useful 
indirect information about the full dynamics of the system from the properties of the DBI solution. In the 
TDBI formulation, where this relation is manifest geometrically, one is instructed to think in terms of a 
single object ---the tachyon-paperclip--- which incorporates both degrees of freedom. This point 
of view implies, in particular, that a modulus of the DBI solution, $e.g.$ the turning point value 
$\rho_0$ in our present example, is necessarily a modulus of the full tachyon-paperclip and 
therefore also a modulus of the \ddbar\ tachyon. On a practical level, this is a way in which the 
DBI action can be used to read off quickly some of the properties of the \ddbar\ tachyon, $e.g.$
that there is a marginal tachyon condensate on the disc worldsheet with a specific winding 
quantum number. As another implication of this relation, the flavor chiral symmetry breaking 
that occurs geometrically by reconnection in the DBI solution is not independent of the \ddbar\ 
tachyon condensation which carries explicitly the order parameter for this breaking.

\section{Flavor D8-branes in the Sakai-Sugimoto model}
\label{ss}

Many of the qualitative features of flavor branes in the NS5-D$p$-$\overline{{\rm D}p}$ 
system are expected to persist also in 
the case of the Sakai-Sugimoto model, $i.e.$ when we replace the NS5-branes with
D4-branes. More specifically, we still expect the open string theory on the D8-branes to 
be controlled by two closely related interactions: an interaction with non-vanishing
winding of the \ddbar\ tachyon and an interaction with vanishing winding of the transverse 
scalars. An additional argument in favor of this picture was given in \cite{Antonyan:2006vw}. 
Both interactions are relevant for the low-energy dynamics of the brane irrespective
of how large the asymptotic separation of the D8 hairpin legs is. Only the latter interaction 
with vanishing winding will be captured efficiently by the DBI action and only in a region 
of the solution where the \ddbar\ tachyon is mildly condensed. Marginal directions in the 
space of solutions of the DBI action imply marginal directions in the coupled system of 
transverse scalars and \ddbar\ tachyon.

To determine the low-energy dynamics of the flavor D8-branes in the Sakai-Sugimoto
model we need an effective action that incorporates the interactions of the \ddbar\ tachyon.
Extending the lessons of the previous sections we would like to propose that the tachyon-DBI 
action for a non-BPS D9-brane is a promising tool for this purpose. The equilibrium 
configurations of the D8-branes can be determined by solving the equations of motion of the 
tachyon-DBI action
\beq
\label{ssaa}
\SS=- \int du\, dx^4 \, u^{13/4}V(T)
\sqrt{1+g(u)\left(\d_u T\right )^2+\frac{1}{g(u)}\left(\d_{x^4} T\right)^2}~, ~~
g(u)=\left(1-\frac{u_{\rm KK}^3}{u^3}\right)\left( \frac{u}{R}\right)^{\frac{3}{2}},
\eeq
or the equations of motion of the corresponding action in the $(\rho,\sigma)$ parametrization of 
$T$ and $x^4$, to find `condensing tachyon-paperclip' solutions with the right asymptotics. 
In the language of the complex-valued \ddbar\ tachyon $\TT$ the asymptotics are controlled by 
two parameters: the parameter of the normalizable branch (with strength $\mu$) and the parameter 
of the non-normalizable branch (with strength $M$). Holographically, $\mu$ controls the 
order parameter of flavor chiral symmetry breaking and $M$ the bare quark mass.
There is a non-trivial transformation that relates $\TT$ with $T$ and $\mu,M$ with the 
coefficients of the leading and subleading branches of $T(u,0)$ evaluated at the maximum or
minimum of the tachyon-paperclip. This transformation is part of the holographic dictionary
when expressed in the language of the tachyon-DBI action.

In previous sections we kept the tachyon potential $V(T)$ free. Once again, we would like to
know what kind of potentials reproduce sensible solutions. At large-$T$
an exponential behavior of the form \eqref{asyaa}
\beq
\label{ssab}
V(T)\sim e^{-\beta T}~, ~~ T\gg 1
~,
\eeq
which is appropriate for the NS5-brane case, has the following feature in the Sakai-Sugimoto 
model. It is simple to verify that in this case the equation of motion following from the variation of the
action \eqref{ssaa} depends only on the tachyon derivatives $\d_u T$ and $\d_{x^4}T$.
Hence, a tachyon solution can be shifted freely by a constant. In periodic solutions of the 
general type of Fig.\ \ref{rollprofile} this free shift occurs without changing the period in the 
angular $x^4$ direction. If it is not possible to fix this free shift by some extra condition,
$e.g.$ a regularity condition, this freedom would imply that there is a 
one-parameter family of tachyon-paperclip solutions with a modulus-independent period as in 
the NS5-brane case. This feature would contradict the information coming from the DBI action, where 
one finds a one-parameter family of hairpin solutions with a modulus-dependent asymptotic separation. 
In that case, one would hope to put further constraints on the precise form of the tachyon potential 
$V(T)$ by making further use of the linked properties of the \ddbar\ tachyon and transverse
scalars. 

The tachyon-DBI formulation also changes the way we compute mesonic spectra in holographic
setups. Typically, in order to determine the structure of the mesonic spectra, we compute fluctuations
of the transverse scalars and gauge fields using the DBI action. In the Sakai-Sugimoto model
we need to add the \ddbar\ tachyon and its fluctuations. In a formalism that treats the transverse
scalar $x^4$, the $U(1)\times U(1)$ gauge fields $A_L$ and $A_R$, and the \ddbar\ tachyon 
$\TT$ as the fundamental degrees of freedom one would find four different sectors of mesons: 
vector mesons (from the fluctuations of $A_+=A_L+A_R$), axial vector mesons (from the 
fluctuations of the transverse part of $A_-=A_L-A_R$), pseudoscalar mesons (from the 
fluctuations of the transverse scalar $x^4$ and the longitudinal part of $A_-$) and
scalar mesons (from the fluctuations of the \ddbar\ tachyon $\TT$) 
\cite{Casero:2007ae,Bergman:2007pm,Dhar:2008um}.

In the TDBI language we are instructed to compute instead fluctuations of a real
tachyon $T(u,\sigma)$, a transverse scalar $x^4(u,\sigma)$ and a $U(1)$ gauge field $A(u,\sigma)$. 
The fluctuations of $T$ and $x^4$ can be viewed as fluctuations of an auxiliary 2-surface (with 
a cigar topology) in the extended three-dimensional space $(T,u,x^4)$. The above four 
mesonic sectors will arise in this language in sectors with different $\sigma$-dependence. 

The TDBI properties of D8-brane dynamics in the Sakai-Sugimoto model will be discussed more 
extensively in \cite{companion}.

\section{Conclusions}
\label{conlusions}

Hairpin-branes appear frequently in holographic contexts that exhibit flavor
chiral symmetry breaking. We argued that the low-energy spectrum of the 
open string theory on these branes includes, besides the standard gauge 
field and transverse scalars, a light complex scalar field. Following the 
setup of Ref.\ \cite{Erkal:2009xq} we explored whether the abelian tachyon-DBI action
can describe the low-energy dynamics on hairpin-branes capturing efficiently the 
non-local dynamics of a bifundamental complex scalar field on them. Our main
purpose in this paper was to test how well this effective field theory performs
in setups where open string theory can be solved with $\alpha'$-exact methods.

We focused on the NS5-D$p$-$\overline{{\rm D}p}$ system in a regime that is 
captured by hairpin D$p$-branes in the near-horizon region of the Wick-rotated 
black NS5-brane. We showed that a radially condensing tachyon-paperclip
solution of the abelian tachyon-DBI action gives a satisfactory description of some of the 
key features of the exact string solution and used the available information from string
theory to put constraints on the tachyon potential $V(T)$ (under a certain assumption for
the tachyon-paperclip solution). These results can be used 
to study further, within a well motivated effective field theory, deformations of the standard
hairpin-branes where the non-normalizable branch of the \ddbar\ tachyon is also
turned on.

A major motivation behind this work has been to understand better similar situations in 
holographic backgrounds with RR fields where an explicit open string description is 
currently out of reach. D8 hairpin-branes in the Sakai-Sugimoto model for QCD is one of 
these cases. One would like to apply the above framework in these more complicated 
situations, establish the corresponding rules of holography and obtain new lessons
for the flavor dynamics of strongly coupled large-$N$ gauge theories. Phenomenologically
interesting quantities like the spectrum of mesons should be computed now under a
new prism \cite{companion}.

Besides the very interesting applications to gauge theory we believe that this exercise 
will also be useful in uncovering new information about open string dynamics in curved 
backgrounds. One would like to test the applicability of the general action \eqref{ddbaraa} in 
diverse situations and derive appropriate constraints on the tachyon potential $V(T)$.

\medskip
\section*{Acknowledgements}
\noindent

I am grateful to Vincenzo Cal\`o for the collaboration during the initial stages of this work, 
and to Emilian Dudas, Elias Kiritsis, David Mateos, Angel Paredes and Ashoke Sen for 
instructive discussions. This work has been supported by the European Union through 
an Individual Marie Curie Intra-European Fellowship. Additional support was provided
by the ANR grant, ANR-05-BLAN-0079-02, the RTN contracts 
MRTN-CT-2004-005104 and MRTN-CT-2004-503369, and the CNRS
PICS \#~  3059, 3747, and 4172.

\section*{Appendices}

\begin{appendix}

\section{Worldsheet boundary interactions for hairpin-branes on the cigar}
\label{cosmo}

In this appendix we review the worldsheet boundary interactions that characterize 
hairpin-branes on the $\NN=(2,2)$ supersymmetric $SL(2)_k/U(1)$ coset following the analysis
of \cite{Hosomichi:2004ph} and derive the first relation in eq.\ \eqref{cftam}.

The hairpin-branes of the supersymmetric coset are related by mirror symmetry to the
B-type branes of $\NN=2$ Liouville theory that were analyzed in \cite{Hosomichi:2004ph}.
The worldsheet boundary interactions characterizing these branes are discussed in 
section 5.1 of that paper. They involve two types of boundary contributions to the worldsheet
action. The first one is
\beq
\label{cosmoaa}
\oint dx\, \left[\overline{\lambda} \d_x \lambda
-(\mu_B \lambda+\mu_{\bar B} \overline{\lambda})e^{-\frac{k}{2}(\rho+i\tilde \theta)+iH}
-(\bar \mu_B\lambda+\bar \mu_{\bar B} \overline{\lambda})e^{-\frac{k}{2}(\rho-i\tilde \theta)-iH}
\right]
\eeq
where $\lambda$, $\overline{\lambda}$ are boundary fermions and 
$\mu_B,\mu_{\bar B},\bar \mu_B,\bar \mu_{\bar B}$ are the corresponding boundary 
couplings and the second one is 
\beq
\label{cosmoab}
-\oint dx\, \tilde \mu (\lambda \overline{\lambda}-\overline{\lambda}\lambda) (\d H-k\d\theta)e^{-\rho}
\eeq
where $\tilde \mu$ is the corresponding boundary coupling.

Using the necessary bootstrap techniques Ref.\ \cite{Hosomichi:2004ph} shows that the boundary
couplings can be expressed in terms of the brane labels $(J,M)$ and bulk coupling $\mu_{bulk}$
in the following way
\beq
\label{cosmoac}
(\mu_B,\mu_{\bar B},\bar \mu_B,\bar \mu_{\bar B})=
i\sqrt{\frac{k\mu_{bulk}}{2\pi}}
\left(e^{i\pi(J-M)},e^{-i\pi(J-M)},e^{-i\pi(J+M)},e^{i\pi(J+M)}\right)
~,
\eeq
\beq
\label{cosmoad}
\tilde \mu=-i\sin\left(\frac{\pi}{k}(2J-1)\right) 
\frac{\mu^{\frac{1}{k}}_{bulk}\Gamma\left(-\frac{1}{k}\right)} {2\pi k}
~.
\eeq
Setting $J=\frac{1}{2}+iP$ and $M=0$ we obtain
\beq
\label{cosmoae}
(\mu_B,\mu_{\bar B},\bar \mu_B,\bar \mu_{\bar B})=
\sqrt{\frac{k\mu_{bulk}}{2\pi}}
\left(-e^{-\pi P},e^{\pi P},e^{\pi P},-e^{-\pi P} \right)
\eeq
\beq
\label{cosmoaf}
\tilde \mu=\frac{\mu^{\frac{1}{k}}_{bulk}\Gamma\left(-\frac{1}{k}\right)} {2\pi k}
\sinh\left(\frac{2\pi P}{k}\right)
~.
\eeq

For non-zero $P$ we can rotate the boundary fermions
\beq
\label{cosmoag}
\mu \xi=\mu_B \lambda+\mu_{\bar B} \overline{\lambda}~, ~~
\bar \mu \overline{\xi}=\bar \mu_B \lambda+\bar \mu_{\bar B} \overline{\lambda}
\eeq
with 
\beq
\label{cosmoai}
\mu\bar \mu=\bar \mu_B \mu_{\bar B}-\mu_B \bar \mu_{\bar B}
\eeq
to recast the boundary interaction \eqref{cosmoaa} into the form
\beq
\label{cosmoaj}
\oint dx\, \left[\overline{\xi} \d_x \xi
-\mu \xi e^{-\frac{k}{2}(\rho+i\tilde \theta)+iH}
-\bar \mu \overline{\xi} e^{-\frac{k}{2}(\rho-i\tilde \theta)-iH}
\right]
~.
\eeq
These relations define the boundary couplings $\mu,\bar{\mu}$ that appear in the
main text (eqs.\ \eqref{cftal}, \eqref{cftam}). Inserting the values \eqref{cosmoae} into
\eqref{cosmoai} we obtain 
\beq
\label{cosmoak}
\mu \bar \mu=\frac{k\mu_{bulk}}{\pi} \sinh(2\pi P)
\eeq
reproducing the first relation in \eqref{cftam}.

Observe that $P=0$ is a special case for this manipulation. In that case,
the transformation $(\lambda,\overline{\lambda})\to (\xi,\overline{\xi})$ \eqref{cosmoag}
is singular. Setting directly $P=0$ in \eqref{cosmoaa} one finds the boundary interaction
\beq
\label{cosmoal}
\oint dx\, \left[\overline{\lambda} \d_x \lambda
-\sqrt{\frac{k\mu_{bulk}}{2\pi}} (- \lambda+ \overline{\lambda})e^{-\frac{k}{2}(\rho+i\tilde \theta)+iH}
+\sqrt{\frac{k\mu_{bulk}}{2\pi}} (-\lambda+ \overline{\lambda})e^{-\frac{k}{2}(\rho-i\tilde \theta)-iH}
\right]
\eeq
which sets the boundary couplings of the vertex operators $e^{-\frac{k}{2}(\rho\pm i\tilde \theta)\pm iH}$
in a different fashion.

\section{Tachyon-paperclips and the \ddbar\ tachyon in general holographic backgrounds}
\label{tachyons}

In this appendix we consider the general setup of a holographic background with a radial 
coordinate $u$ and an angular coordinate $\theta$. We restrict attention to a diagonal background 
metric that depends only on $u$ and analyze the TDBI action for a non-BPS D$(p+1)$-brane 
oriented along a set of directions $x^\mu$ $(\mu=0,1,\ldots, p-1)$, that exhibit Lorentz invariance,
and $u,\theta$. The tachyon potential $V(T)$ is chosen to have the generic form of Fig.\ 
\ref{V(T)}. We assume that the corresponding equations of motion have a tachyon-paperclip solution 
$T=T(u,\theta;\phi)$ at large $u$ that depends on a free parameter $\phi$, $i.e.$ $\phi$ is a 
modulus of the solution. 

Let us promote $\phi$ to a slowly-varying field that depends on the coordinates $x^\mu$
\beq
\label{TTaa}
\phi=\phi(x^\mu)
~.
\eeq
Inserting the asymptotic solution $T(u,\theta;\phi)\equiv T_\phi$ into the TDBI action we expand 
up to quadratic order in derivatives $\d_\mu \phi$. Integrating out the $u,\theta$ dependence 
and redefining $\phi$ to a canonically normalized field $\phi_{can}$ we obtain the kinetic term
\beq
\label{TTab}
\SS_{kin}\sim \int dx^0\cdots dx^{p-1}\, (\d_\mu \phi_{can})^2
\eeq
and no potential terms. The absence of potential terms is a consequence of the assumption 
that $\phi$ is a modulus of the asymptotic paperclip solution.

According to the discussion in the main text the paperclip solution captures implicitly the 
physics a hairpin D$p$-brane oriented along the spacetime directions $x^\mu$, $u$. 
In this appendix we are interested in identifying the relation between the hairpin-brane
\ddbar\ tachyon $\TT$ and the TDBI tachyon $T$.

For large positive $T$ and real $\TT$ we postulate the following relation between $\TT$ and 
$\phi_{can}$
\beq
\label{TTac}
\TT\sim \phi_{can}^\zeta
~.
\eeq
$\zeta$ is a constant related to the $U(1)_L-U(1)_R$ charge of $\phi_{can}$.
$\TT$ is implicitly defined here with an appropriate $u$-dependent normalization factor
that cancels its leading $u$-dependence.
In flat space \cite{Erkal:2009xq} there is no $u$-dependence and
\beq
\label{TTad}
\phi_{can}\sim e^{-\frac{T}{2\sqrt 2}}~, ~~ \zeta=2
~.
\eeq

To identify the kinetic term \eqref{TTab} we insert the asymptotic solution $T_\phi$ into 
the TDBI action and expand to quadratic order. The TDBI action reads
\beq
\label{TTae}
\SS=-\int dx^0\cdots dx^{p-1} dud\theta\, V(T) \sqrt{\det A}
\eeq
with background metric
\beq
\label{TTaf}
ds^2=dT^2+g_{\mu\nu}(u) dx^\mu dx^\nu+g_{uu}(u) du^2+g_{\theta\theta}(u) d\theta^2
\eeq
and 
\begin{subequations}
\bea
\label{TTag}
A_{\mu\nu}&=&g_{\mu\nu}+(\d_\phi T)^2 \d_\mu \phi \d_\nu \phi
~,
\\
A_{\mu u}&=&\d_\phi T \d_u T \d_\mu \phi~~, ~~
A_{\mu \theta}=\d_\phi T \d_\theta T \d_\mu \phi
~,
\\
A_{uu}&=&g_{uu}+(\d_u T)^2~, ~~
A_{\theta\theta}=g_{\theta \theta}+(\d_\theta T)^2~, ~~
A_{u\theta}=\d_u T \d_\theta T
~.
\eea
\end{subequations}
By assumption
\beq
\label{TTaga}
g_{\mu\nu}(u)=g(u)\eta_{\mu\nu}
~.
\eeq

Let us denote by $\AA$ the $p\times p$ matrix with elements $A_{\mu\nu}$, by 
$\BB$ the $p\times 2$ matrix $(A_{\mu u},A_{\mu \theta})$, by $\BB^T$ its transverse 
and by $\CC$ the $2\times 2$ matrix 
\beq
\label{TTai}
\left(
\begin{array}{c c}
A_{uu} & A_{u\theta} \\
A_{u\theta} & A_{\theta \theta} 
\end{array}
\right)
~.
\eeq
A useful identity for the computation of $\det A$ is
\beq
\label{TTaj}
\det \left(
\begin{array}{c c}
\AA & \BB \\
\BB^T & \CC 
\end{array}
\right)
=\det \AA \cdot \det \left(\CC-\BB^T \AA^{-1} \BB\right)
~.
\eeq
After some algebra we find
\beq
\label{TTak}
\det A=g \det \CC-\left( \det \CC-g_{uu}g_{\theta\theta}\right) (\d_\phi T) ^2 (\d_\mu \phi) ^2
~.
\eeq
Expanding the square root
\beq
\label{TTal}
\sqrt{\det A}\simeq \sqrt{g \det \CC}
-\frac{1}{2}\frac{\det \CC-g_{uu}g_{\theta\theta}}{\sqrt{ g \det \CC}} (\d_\phi T) ^2 (\d_\mu \phi) ^2
+\ldots
~.
\eeq
The diverging factor
\beq
\label{TTam}
\KK(\phi)=\int dud\theta\, V(T_\phi) 
\frac{\det \CC-g_{uu}g_{\theta\theta}}{\sqrt{ g \det \CC}} (\d_\phi T_\phi) ^2
\eeq
is a function of $\phi$ whose computation requires explicit information about the tachyon 
potential and the tachyon-paperclip solution. In terms of $\KK(\phi)$ the kinetic term \eqref{TTab} 
becomes
\beq
\label{TTan}
\SS_{kin}\sim -\frac{1}{2} \int dx^0\cdots dx^{p-1}\, \KK(\phi) (\d_\mu \phi)^2
\eeq
from which we can read off $\phi_{can}$.

For the tachyon-paperclip solution of the NS5-D$p$-$\overline{{\rm D}p}$ system
\eqref{asyab} we have (renaming $u$ as $\rho$)
\beq
\label{TTao}
g=1~, ~~ T_\phi=\phi+b\rho+c\log \cos\theta~, ~~
~~ \det \CC=\frac{kc^2}{\cos^2 \theta}~,
~~ V(T)=e^{-\beta T}
~.
\eeq
Hence, up to a $\phi$-independent diverging factor that comes from the $\rho,\theta$ integration
\beq
\label{TTap}
\KK(\phi)\sim e^{-\beta \phi}
\eeq
and
\beq
\label{TTaq}
\phi_{can}\sim e^{-\frac{\beta}{2} \phi}
~,
\eeq
which together with \eqref{TTac} implies the transformation \eqref{asyak} of section \ref{main}.

\section{Further numerical results on the relation $\mu_T(\rho_0)$: the case of generic $k$}
\label{extranumerics}

\noindent
{\it Note added}: This appendix has been added to the second arXiv version of the paper
in September of 2010 and does not appear in the published Nuclear Physics B version 
which appeared earlier.

\

Following the logic outlined in the beginning of section \ref{numeric}, we consider here a 
differential equation that provides an estimate to the behavior of the tachyon maximum
$T(\rho,0)$ for generic $k$. We assume that $T\gg 1$ and approximate the tachyon potential
by the large-$T$ asymptotics \eqref{asyaa}. It turns out that this approximation does not affect 
the final result significantly ---a direct comparison to the results obtained in section \ref{numeric} 
for $k=2$ with the full potential $\cosh^{-1}\left(T/\sqrt 2 \right)$ reveals small numerical 
differences.

In what follows we insert the ansatz 
\beq
\label{enumaa}
T(\rho,\theta)=f(\rho)+c \log\cos\theta
\eeq
into the equation of motion \eqref{tdbiad}, expand to leading order around $\theta=0$ 
and obtain the differential equation
\beq
\label{enumab}
\left( f' \cosh\rho - \left( \frac{c \cosh^2\rho}{\sinh\rho}-k\beta \sinh\rho \right) \right)
\left( k+{f'}^2\right) +k \sinh\rho f''=0
~.
\eeq
We are leaving $\beta$ as a free constant and express $c$ in terms of $\beta$ using 
the equations \eqref{asyad}, \eqref{asyae}
\beq
\label{enumac}
c=\frac{1}{2}\left( \frac{1}{\beta}+\beta k \right)
~.
\eeq
In the special case of $\beta=\frac{1}{\sqrt k}$ (see eq.\ \eqref{asyar}) the equation 
\eqref{enumab} becomes
\beq
\label{enumad}
\left( f' \cosh\rho - \frac{\sqrt k}{\sinh\rho} \right)
\left( k+{f'}^2\right) +k \sinh\rho f''=0
\eeq
which reduces correctly to eq.\ \eqref{numab} for $k=2$ and $f\gg 1$.

We can solve numerically the differential equation \eqref{enumab} for different values of 
$k$ and $\beta$. For each of these values we obtain a curve that expresses, at fixed and 
large $M_T$,  the relation between the boundary value $\mu_T$ and the singularity 
point $\rho_*$ (which is treated here as an approximation to the turning point of the brane).
Analytically this curve is expected to be of the form \eqref{numaca}, \eqref{numacaab}.
We find that the numerically determined curve does not obey this functional relation 
for generic values of $\beta$.\footnote{For generic values of $\beta$ there is a shift of
the point where $\mu_T$ diverges away from the tip of the cigar.}
A qualitative and quantitative match to the expected analytical result is observed only 
when $\beta$ takes the special value $\frac{1}{\sqrt k}$ \eqref{asyar}. We view this result 
as strong numerical evidence in favor of the expectation expressed below eq.\ \eqref{asyar} that 
the expressions in \eqref{asyar} provide the correct tachyon asymptotics for all values of $k$.

\begin{figure}[t!]
\centering
\includegraphics[height=6cm]{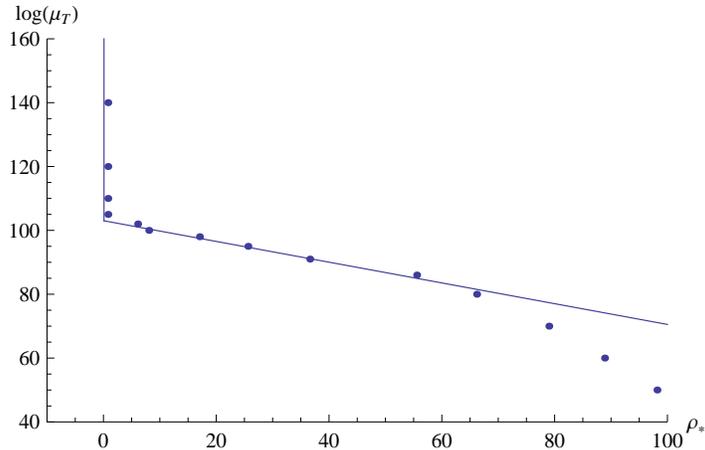}
\bf\caption{\it \small Fourteen numerically determined values of $\log(\mu_T)$ versus
$\rho_*$ are plotted in this figure by evaluating the solution of eq.\ \eqref{enumad} for 
$k=10$ and $M_T=e^{150}$. The solid curve is a fit based on the functional form 
\eqref{numaca} with $y=6.5$ and $z=0.05$.}
\label{muT}
\end{figure}

Fig.\ \ref{muT} depicts the numerically determined curve $\log(\mu_T)$ as a function of
$\rho_*$ for a randomly chosen value of $k$ ($k=10$), the special value of $\beta$
($\beta=\frac{1}{\sqrt {10}}$), and $M_T=e^{150}$. The solid curve is a fit based on the 
analytically expected form \eqref{numaca} with fit values $y=6.5$ and $z=0.05$. The analytic
values of $y$ and $z$ compare well with the fit values ---for $k=10$ these are 
(see eq.\ \eqref{numacaab}) $y=k=10$ and $z=\frac{1}{18}\simeq 0.0556$. 
As in section \ref{numeric} we observe an expected mismatch between the fit and the 
numerically determined points in the vicinity of the point where the boundary conditions 
are placed (here $\rho=100$).

Similar features are observed for generic values of $k$.
The agreement between the fit value and the analytical value of the slope of the linear section of the 
curve, $i.e.$ the parameter $yz$, appears to improve as we increase $k$. For $k=2$ 
(see e.g.\ section \ref{numeric}) we obtain $(yz)_{num} \simeq 0.4$ versus $(yz)_{anal}=1$. 
For $k=10$ we obtain $(yz)_{num}\simeq 0.325$ versus $(yz)_{anal}=\frac{10}{18}\simeq 0.556$.

\end{appendix}


\end{document}